\documentclass[12pt,a4paper]{article}
\usepackage{amsmath}
\usepackage[T1]{fontenc}
\usepackage[utf8]{inputenc}
\usepackage{authblk}
\usepackage{slashed}

\usepackage{hyperref}
\usepackage{graphicx}
\usepackage{fancybox}
\usepackage{color}
\usepackage{epsfig,graphicx}
\usepackage{amsfonts}

\def\n{\nu}

\title{A Holographic Quantum Hall Ferromagnet}

 \author[1]{C.\ Kristjansen}
\author[2,3]{R.\ Pourhasan}
\author[4,5]{G.W.\ Semenoff}

\affil[1]{Niels Bohr Institute, Copenhagen University, Blegdamsvej
17,   2100 Copenhagen \O, Denmark} \affil[2]{Perimeter Institute
for Theoretical Physics, 31 Caroline St. N., Waterloo, ON, N2L
2Y5, Canada} \affil[3]{Department of Physics \& Astronomy,
University of Waterloo, Waterloo, Ontario N2L 3G1, Canada}
\affil[4]{Department of Physics and Astronomy, University of
British Columbia,  Vancouver, BC Canada V6T 1Z1}
\affil[5]{International Institute of Physics, Federal University of Rio Grande do Norte, Av.~Odilon Gomes
de Lima 1722, Capim Macio, Natal-RN 59078-400 Brazil}

\date{} 
\begin{document}

\maketitle

\begin{abstract}
 A detailed numerical study of a recent proposal for exotic states of the D3-probe D5 brane system with charge density and
an external magnetic field is presented.   The state has a large number of coincident D5 branes
blowing up  to a D7 brane in the presence of
the worldvolume electric and magnetic fields which are necessary to construct the holographic state.
Numerical solutions have shown that these states can
compete with the the  previously known chiral symmetry breaking and maximally symmetric phases
of the D3-D5 system.  Moreover, at integer filling fractions, they are incompressible with integer
quantized Hall conductivities.   In the dual superconformal defect field theory, these solutions correspond
to states which break the chiral and global flavor symmetries spontaneously.
The region of the temperature-density plane where the D7 brane has lower
energy than the other known D5 brane solutions is identified.  A hypothesis for the structure of states with
filling fraction and Hall conductivity greater than one is made and tested by numerical computation.
A parallel with the quantum Hall ferromagnetism or magnetic catalysis phenomenon
which is observed in graphene is drawn.  As well as demonstrating that the phenomenon can
exist in a strongly coupled system, this work makes a number of predictions of symmetry breaking
patterns and phase transitions for such systems.
\end{abstract}

\newpage
\section{Introduction and Summary}

The quantum Hall effect is one of the most dramatic phenomena in condensed matter physics
\cite{qhe1} \cite{qhe2}.
At particular values of its charge
density and
magnetic field, a two-dimensional electron gas exhibits incompressible charge-gapped states.
These states can be robust
and persist over a range of the ratio of density to field, that is, over a Hall plateau on which the Hall conductivity is a constant $\nu$
times the elementary unit of conductivity $\frac{e^2}{h}$ ($=\frac{1}{2\pi}$ in the natural units which we shall use in this paper),
\begin{align}
\sigma_{xy}=\frac{\nu}{2\pi}.
\label{hallconductivity}
\end{align}
 The same electron gas can exhibit an array of such states, where $\nu$ is generally
an integer, for the integer quantum Hall effect, or a rational number for the fractional quantum Hall effect.

What is more, the integer quantum Hall effect has a beautiful and simple explanation as a single-particle
phenomenon.  When a charged particle moving in two dimensions is exposed to a magnetic field, its spectrum is resolved into
discrete Landau levels. Landau levels are flat, dispersionless bands with gaps between them.
Fermi-Dirac statistics dictates that the low energy states of a many-electron system are obtained by filling
the lowest energy single-electron states, with one electron per state.  When a Landau level is completely filled with electrons,
the next electron one inserts into
the system must go to the next higher level which is separated from the ones that are already occupied by a gap.
The result is a jump in the chemical potential.  Alternatively, when the chemical potential
is in the gap between levels, it can be varied with no change of the charge density.  Such a state is said to be incompressible.
This effect is enhanced by disorder induced localization which forms a mobility gap and results in the Hall plateau.

In the absence of
disorder, for free electrons, the Hall conductivity is given by (\ref{hallconductivity}) with $\nu$ equal to the filling fraction.  The filling fraction
is defined as the ratio of the number of electrons to the number of states in the Landau levels which are either completely or partially occupied
(see (\ref{fillingfraction}) below).  When a number of Landau levels are completely filled,   $\nu$ is an integer which coincides with the number
of filled levels and the Hall conductivity is quantized.  It is given by the formula (\ref{hallconductivity}).
Moreover,  for completely filled energy bands, the Hall conductivity is a topological quantum number
insensitive to smooth alterations of the energy band \cite{laughlin}-\cite{thouless2},
such as those caused by changes in the environment of the single electrons.
We can turn on
lattice effects and disorder with the Hall conductivity remaining unchanged,
and can thus conclude that, when $\nu$ is an integer, the quantized
Hall conductivity is robust for a large range of single-particle interactions including the effects of disorder which
are responsible for forming the Hall plateaus.

In addition to this,
there are good theoretical arguments for the persistence of the integer
Hall effect in the presence of electron-electron interactions, at least
when the interactions are weak enough that perturbation theory can be applied.
An easy way to understand this is by noting that, at the
level of a low energy effective action,
the Hall effect is encoded in a Chern-Simons term for the photon field,
\begin{align}\label{chernsimons}
S_{\rm CS}=\frac{\sigma_{xy}}{2}\int d^3x\,\epsilon^{\mu\nu\lambda}A_\mu \partial_\nu A_\lambda
~.\end{align}
The coefficient of the Chern-Simons term is proportional to the Hall conductivity.
Moreover, there is a theorem which states that the Chern-Simons term does not renormalize beyond one-loop order in either a
relativistic or non-relativistic field theory \cite{Coleman:1985zi}
\cite{Semenoff:1988ep} \cite{Lykken:1991xs}.   The theorem depends on the existence of a charge gap in the spectrum.  If the gap
closes it is known that either scalar or fermion charged matter can renormalize the
Chern-Simons term \cite{Semenoff:1988ep}.
Thus, as far as perturbation theory is valid, the existence of the integer quantum Hall effect after  interactions are turned on
is intimately tied to the question of whether the incompressible nature of the state due to the
energy gap between Landau levels survives in the interacting theory.

In this paper, we shall discuss the question as to whether any features of the integer
Hall effect can persist when the coupling is strong, beyond the reach of perturbation theory.
The development of AdS/CFT duality between certain gauge field
theories and certain string theories has given us a tool for solving the strong coupling limit of some quantum systems.
In a recent paper \cite{Kristjansen:2012ny}, two of the authors have found an example of a strongly coupled quantum field theory which, in
a state with non-zero charge density and when
subject to an external magnetic field,  exhibits incompressible
states with integer quantized Hall conductivity.  It occurs in a non-Abelian gauge field theory that has a well-established
string theory dual, the D3-D5 system
\cite{Karch:2000gx}-\cite{Karch:2000ct}. The string theory is quantitatively tractable in its semi-classical low energy limit, and a further
probe limit where the number of D5 branes is much smaller than the number of D3 branes.  These limits
coincide with the strong coupling and quenched planar limit of the quantum field theory
and its solution yields information about the latter at strong coupling.
The behavior of the theory when a charge density and magnetic field are added can readily be studied there.  In that system, it was shown that
there exist
exotic states of the D3-D5 system where it becomes incompressible.  These states occur at precisely integer values of the filling fraction $\nu$,
where
\begin{align}
\nu\equiv \frac{2\pi\rho}{NB},
\label{fillingfraction}
\end{align}
 with $\rho$ the particle density, $B$  the external field and $N$ the number of colors of quarks in the non-Abelian gauge theory.
These states were argued to be the natural strong coupling manifestation of some incompressible
integer quantum Hall states which appear in the weak coupling limit of that theory.

\begin{figure}
\begin{center}
\scalebox{0.7}{
\begin{picture}(250,250)(0,0)
\put(0,100){\vector(1,0){220}}
\put(100,0){\vector(0,1){220}}
\thicklines
\put(104,120){\line(1,0){22}}
\put(104,116){\oval(8,8)[tl]}
\put(126,124){\oval(8,8)[br]}
\put(134,160){\line(1,0){22}}
\put(134,156){\oval(8,8)[tl]}
\put(156,164){\oval(8,8)[br]}
\put(164,200){\line(1,0){22}}
\put(164,196){\oval(8,8)[tl]}
\put(74,80){\line(1,0){22}}
\put(74,76){\oval(8,8)[tl]}
\put(96,84){\oval(8,8)[br]}
\put(44,40){\line(1,0){22}}
\put(44,36){\oval(8,8)[tl]}
\put(66,44){\oval(8,8)[br]}
\put(100,84){\line(0,1){32}}
\put(130,124){\line(0,1){32}}
\put(160,164){\line(0,1){32}}
\put(70,44){\line(0,1){32}}
\put(40,4){\line(0,1){32}}
\put(89,116,){$\frac{1}{2}$}
\put(89,156,){$\frac{3}{2}$}
\put(105,76,){$-\frac{1}{2}$}
\put(105,36,){$-\frac{3}{2}$}
\put(105,220,){$\sigma_{xy}$}
\put(210,105,){$\rho$}
\end{picture}

}
\end{center}
\caption{\label{figure001} Integer quantum Hall effect in graphene.  The vertical axis is the Hall conductivity in units of  $4\frac{e^2}{h}$.
The horizontal axis is the charge density at fixed magnetic field.   The plateaus occur at the anomalous integer Hall conductivities \mbox{$\sigma_{xy}=4\,\frac{e^2}{h}(n+\frac{1}{2})$.} }
\end{figure}
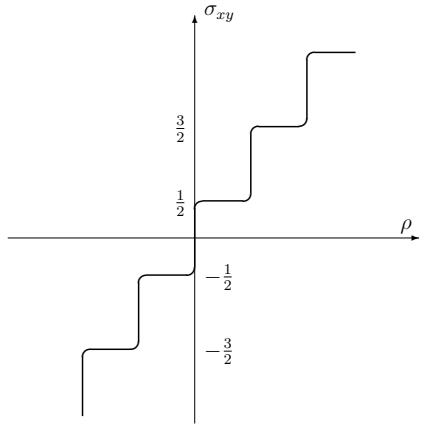

Aside from a manifestation of an integer quantum Hall state, the incompressible states of the strongly coupled system found in
reference \cite{Kristjansen:2012ny} have
an interesting analog in the observed quantum Hall states of graphene.   Graphene is a two-dimensional semi-metal where the
electron obeys an emergent massless Dirac equation with four flavors of the fermion field and an effective SU(4) symmetry \cite{Semenoff:1984dq}
\cite{geim}.
This fact leads to the anomalous quantum Hall effect \cite{Gusynin:2005pk} where the Hall conductance is quantized as
\begin{align}
\sigma_{xy}^{\rm graphene} = \frac{e^2}{h}\cdot 4\cdot\left(n+\frac{1}{2}\right).
\end{align}
The factor of 4 in this expression arises from the four-fold degeneracy of the low energy fermions and the offset
of $1/2$ in $n+1/2$ is a result of the spectrum of the massless Dirac fermions in a magnetic field.  The Dirac Hamiltonian
in a magnetic field has zero energy modes which form a Landau level   at the apex of the Dirac cones.
Particle-hole symmetry dictates that,
in the ground state of the many-electron system, when it is charge neutral, as well
as all of the negative energy single electron states of the Dirac Hamiltonian, half of the zero modes should be filled .
The neutral state
is therefore not a Hall plateau.  It has a half-filled Landau level and it
is in the middle of a Hall step, as depicted in figure \ref{figure001}.  These features were
observed in the initial experiments which were performed soon after graphene was discovered in 2004 \cite{qhegraphene1} \cite{qhegraphene2}.

\begin{figure}
\begin{center}
\scalebox{0.7}{
\begin{picture}(500,250)(0,0){

\begin{picture}(250,250)(0,0)
\put(20,120){\vector(1,0){190}}
\put(115,0){\vector(0,1){220}}
\thicklines
\put(119,200){\line(1,0){41}}
\put(115,196){\line(0,-1){152}}
\put(119,196){\oval(8,8)[tl]}
\put(111,40){\line(-1,0){41}}
\put(111,44){\oval(8,8)[br]}
\put(104,156,){$1$}
\put(104,196,){$2$}
\put(120,76,){$-1$}
\put(120,36,){$-2$}
\put(122,215){$\sigma_{xy}$}
\put(200,105,){$\rho$}
\put(232,115){$\longrightarrow$}
\end{picture}
\begin{picture}(250,250)(0,0)
\put(20,120){\vector(1,0){190}}
\put(115,0){\vector(0,1){220}}
\thicklines
\put(104,120){\line(1,0){22}}
\put(104,116){\oval(8,8)[tl]}
\put(126,124){\oval(8,8)[br]}
\put(134,160){\line(1,0){22}}
\put(134,156){\oval(8,8)[tl]}
\put(156,164){\oval(8,8)[br]}
\put(164,200){\line(1,0){22}}
\put(164,196){\oval(8,8)[tl]}
\put(74,80){\line(1,0){22}}
\put(74,76){\oval(8,8)[tl]}
\put(96,84){\oval(8,8)[br]}
\put(44,40){\line(1,0){22}}
\put(66,44){\oval(8,8)[br]}
\put(100,84){\line(0,1){32}}
\put(130,124){\line(0,1){32}}
\put(160,164){\line(0,1){32}}
\put(70,44){\line(0,1){32}}
\put(104,156,){$1$}
\put(104,196,){$2$}
\put(120,76,){$-1$}
\put(120,36,){$-2$}
\put(122,215,){$\sigma_{xy}$}
\put(200,105,){$\rho$}
\end{picture}

}
\end{picture}}
\end{center}
\caption{\label{figure002} Quantum Hall Ferromagnetism/Magnetic Catalysis of chiral symmetry breaking
in graphene.  The four-fold degeneracy of all Landau levels is seen to be completely resolved in experiments
with sufficiently clean samples
with strong enough magnetic fields \cite{newhallplateaus}.  The vertical axis is the Hall conductivity in units of $\frac{e^2}{h}$.  The
horizontal axis is charge density at fixed magnetic field. }
\end{figure}
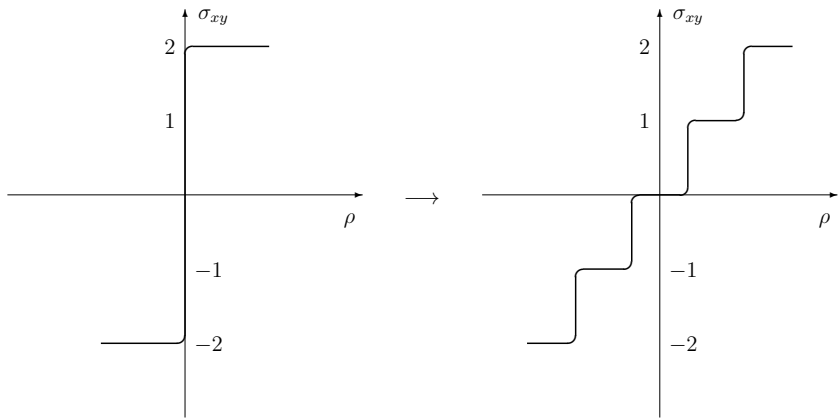

Later, with stronger magnetic fields and cleaner samples, the formation of
new plateaus were discovered at all of the integer steps.  The four-fold degeneracy of the Landau levels is partially
or completely lifted, depending on the filling fraction \cite{newhallplateaus} \cite{kim}.
The new plateaus are depicted in figure \ref{figure002}.
In particular, the states which originate from the zero mode Landau level
have  interesting properties \cite{newhallplateaus1}-\cite{newhallplateaus3}.

The mechanism for formation of these additional plateaus is thought to be spontaneous symmetry breaking and,
like the integer Hall effect itself, it has a beautiful and elegant explanation at weak coupling.
Here, we will focus on the charge neutral Landau level (near $\rho$=0 in figures \ref{figure001} and \ref{figure002}).  As we have already noted,
in an external magnetic field, the single electron spectrum has zero modes and, when we construct the many-electron state
in a certain range of densities, these single-electron zero modes must be partially filled.
If electrons are non-interacting, a partially filled Landau level is a highly degenerate state, as any partial filling has the same
energy as any other partial filling.
In this circumstance, an interaction, no matter how weak, will generically split the degeneracy of these states. The most important electron-electron interaction in graphene is the Coulomb interaction.   There are good arguments that suggest that, for the  quarter-, half-  or three-quarters
filled zero mode Landau level, the
Coulomb  exchange interaction is minimized by states which resolve the four-fold degeneracy by spontaneously breaking the SU(4) symmetry. Once the symmetry is broken,  energy gaps and Hall plateaus emerge at all integer filling fractions.  This phenomenon is known in the condensed matter literature as quantum Hall ferromagnetism \cite{quantumhallferromagnet0}-\cite{quantumhallferromagnet6} and in the particle physics literature as magnetic catalysis of chiral symmetry breaking \cite{cat0}-\cite{Filev:2012ch}.\footnote{These mechanisms usually focus on different order parameters and
are sometimes thought to be mutually exclusive.  In the present case, they are indistinguishable as a nonzero value of one order parameter will
lead to a nonzero value of the other, and in fact the order parameters are equal in the weak coupling limit \cite{Semenoff:2011ya}.}
 In experiments, this resolution is now seen for all of the integer Hall states \cite{kim} as well as the zero modes.
In a clean system, the argument for symmetry breaking that we have reviewed here works at arbitrarily weak coupling and gives a candidate for an 
explanation of this interesting phenomenon.

However,  graphene is not weakly coupled.   The Coulomb interaction in graphene is putatively
strong.\footnote{The Coulomb energy of an electron-hole
pair on neighboring sites is approximately 10 eV, whereas the tunnelling energy between the sites is about 2.7 eV. This is in line with the rough argument
that the graphene fine structure constant which controls the quantum fluctuations of the photon is large: $\alpha_{\rm graphene}= \frac{e^2}{4\pi\hbar v_F}=\frac{e^2}{4\pi\hbar c}\frac{c}{v_F}\approx \frac{300}{137}$ where we have used graphene's emergent speed of light which is a factor of 300 less
than the speed of light in vacuum,  $v_F\approx c/300$.  This suggests that the Coulomb interaction in graphene is strongly coupled and
out of the range of   perturbation theory.  Other indications such as the approximate perfect conical shape of Dirac cones seen in ARPES measurements \cite{Lanzara} suggest that graphene dynamics is approximately scale invariant and has 2+1-dimensional Lorentz invariance over a significant   range of energy scales.  There is no known truly quantitative mechanism which would explain this.} In fact, the magnitude of the energy gaps due to symmetry breaking
that are seen in experiments is of order the
Coulomb energy and they are already large enough to conclude that the system is strongly coupled.
At a first pass, to understand the occurrence of
this symmetry breaking in graphene, it is necessary to understand whether it can also happen in
a strongly coupled system, that is, whether the features of quantum Hall ferromagnetism survive as the coupling
constant is increased to large values.  Reference  \cite{Kristjansen:2012ny} gives an affirmative answer to this question
in the context of a certain quantum field theory.  To be precise, the
supersymmetric large N gauge field theory that is considered there cannot be regarded as a model of graphene
in all of its details.   On the other hand, as we shall outline below, it may be entirely
possible that it does  model some of the physics of the charge neutral Landau level in graphene.
For this reason, among
others, it is
important to have an improved picture of the predictions of the model.

In this paper, we shall present a significant elaboration on the work in reference
 \cite{Kristjansen:2012ny}.
That work considered the D3-D5 system which is dual to a superconformal defect quantum field theory which has
${\cal N}=4$ supersymmetric Yang-Mills theory living in the bulk of 3+1-dimensional spacetime.  The 3+1-dimenisonal
spacetime is bisected by an infinite, flat 2+1-dimensional defect.   A 2+1-dimensional
hypermultiplet field theory resides on the defect and interacts
with the ${\cal N}=4$ degrees of freedom in the 3+1-dimensional bulk.   The defect field theory preserves half of
the supersymmetry of the bulk ${\cal N}=4$ theory  and is conformally symmetric for all values of its coupling constant.
The field theory living on the defect has both scalar and spinor fields and the Lagrangian  is known
explicitly \cite{DeWolfe:2001pq}\cite{Erdmenger:2002ex}.  At weak coupling, its action contains massless fermions and bosons,
\begin{align}\label{weakcouplingaction}
{\cal S}\sim \int d^3x \sum_{\alpha=1}^N\sum_{\gamma=1}^{N_5}\left[\bar\psi_a^{\alpha\gamma} i\slashed\partial\psi_{\alpha\gamma}^a - \partial_\mu\bar\phi_{\dot a}^{\alpha\gamma}\partial^\mu\phi^{\dot a}_{\alpha\gamma}\right]
+\ldots,
\end{align}
which are fundamental representations
of the  gauged $SU(N)$ color and the  global $U(N_5)$ flavor  symmetries and
are $(0,\frac{1}{2})$ and $(\frac{1}{2},0)$ representations of an $SO(3)\times SO(3)$ R-symmetry, respectively.
The masslessness
of the fermions is protected by symmetry as there are no possible  time reversal invariant mass operators which preserve all of
the $SU(N)$, $U(N_5)$ and $SO(3)\times SO(3)$ symmetries.   All solutions that we consider are invariant
under color $SU(N)$, the $U(1)$ subgroup of $U(N_5)$ and the first $SO(3)$.  We will consider solutions which break
either the second $SO(3)$ or the $SU(N_5)$ subgroup of $U(N_5)$ or both, and we will call this ``chiral symmetry breaking'' .  This terminology  will apply to any
solution where the constant $c_2$ defined in equation (\ref{asymptoticpsi}) in section 3 is nonzero. This constant is called the ``chiral condensate''.
  The three dots in (\ref{weakcouplingaction}) indicate the action of ${\cal N}=4$ Yang-Mills theory and interaction terms.

A constant $U(1)\subset U(N_5)$
external magnetic field, $B$, breaks supersymmetry.
The free fermions and free bosons have different Landau level
energies, $\omega_n=\pm\sqrt{2nB}$ and  $E_n=\pm\sqrt{(2n+1)B}$, respectively, with $n=0,1,2,...$.
The boson energies have a gap, $\sqrt{B}$.   At energies lower than this gap, only the bosonic vacuum is relevant.
On the other hand, the fermions have zero modes with degeneracy $2NN_5\frac{B}{2\pi}$.
For states with filling fraction $\nu\leq N_5$ (with $\nu$ defined
in equation (\ref{fillingfraction})
\footnote{Note that there is a factor of the number of colors, $N$, in the denominator of that equation.  We are assuming that the SU(N) color
symmetry remains unbroken.  Thus, we take for candidate states only those which are singlets of the global color symmetry.  For a many-body state of quarks to be a color singlet, the number of quarks must be an integer multiple of $N$.   Therefore we consider states where the quarks come in multiples
of $N$ only.  The filling fraction defined in (\ref{fillingfraction}) is the fractional occupancy of a Landau level where this natural N-fold
degeneracy is taken into account. }),
the lowest energy modes of the sufficiently
weakly interacting theory are   governed by the problem of populating the fermion zero modes.  The arguments for
quantum Hall ferromagnetism should apply to the ${\cal N}=4$ gluon-mediated color interaction.  In the large $N$ planar limit
in particular, the exchange interaction is emphasized and minimizing it should lead to   breaking of the $SO(3)$ chiral symmetry and,
depending on the filling fraction, various symmetry breaking patterns for the $U(N_5)$ flavor symmetry. The states with integer filling
fractions $\nu=0,\pm1,\pm2,...,\pm N_5$ should be gapped, incompressible states with integer quantized Hall conductivities,
though the series could truncate before it gets to $\pm N_5$ if the splitting of the states begins to compete with the  energy of bosons,
which begins at $\sqrt{B}$.\footnote{In addition, once supersymmetry  and scale symmetry are broken by
introduction of a magnetic field, there is no symmetry which
prevents a boson mass term $m^2\bar\phi_{\dot a}^{\alpha\gamma}\phi^{\dot a}_{\alpha\gamma}$ from appearing in effective field theory.
This would further isolate the fermion zero modes.}  Also, the higher fermionic Landau levels are at energies greater than the threshold for creating
 bosons, so one would expect that they lead to no further incompressible states.  We will see shortly that the counting of possible
incompressible states is matched on the strong coupling side which is described by string theory.

The strongly coupled system is described by the embedding of $N_5$ coincident D5-branes in the $AdS_5\times S^5$ background.
At zero charge density and in the absence of magnetic fields, the D5-brane geometry is $AdS_4\times S^2$,
which has superconformal symmetry. The $AdS_4$ is a subspace of $AdS_5$.  The $S^2$ is the maximal volume $S^2$ which can be embedded in $S^5$.
The embedding position has an $SO(3)$ symmetry so that an $SO(3)\times SO(3)$ subgroup of the $SO(6)$ symmetry of $S^5$ survives.

Here, and everywhere in the following, we are considering boundary conditions for the embedding problem
which do not violate the chiral symmetry
of the gauge theory.  This means that, even for other solutions, the worldvolume geometry must approach this maximally
symmetric $AdS_4\times S^2$ geometry sufficiently rapidly as the D5 brane worldvolume approaches the boundary of $AdS_5\times S^5$.
In particular, the $S^2$ must become the $SO(3)\times SO(3)$ symmetric maximal $S^2$
and the $N_5$ multiple D5 branes must become coincident.  The latter condition makes their boundary condition symmetric under $SU(N_5)$.
 If the chiral symmetry is broken, it must be by spontaneous symmetry breaking. This symmetry breaking
occurs if, as the D5 brane worldvolume stretches into the bulk of $AdS_5\times S^5$, either the $S^2$ which the D5 brane wraps deviates at all from the maximal  one of the most symmetric embedding, or the D5
branes spread apart, breaking the global $SU(N_5)$ symmetry.  We will encounter both of these behaviors shortly.

\begin{figure}
\begin{center}
\includegraphics[scale=.65]{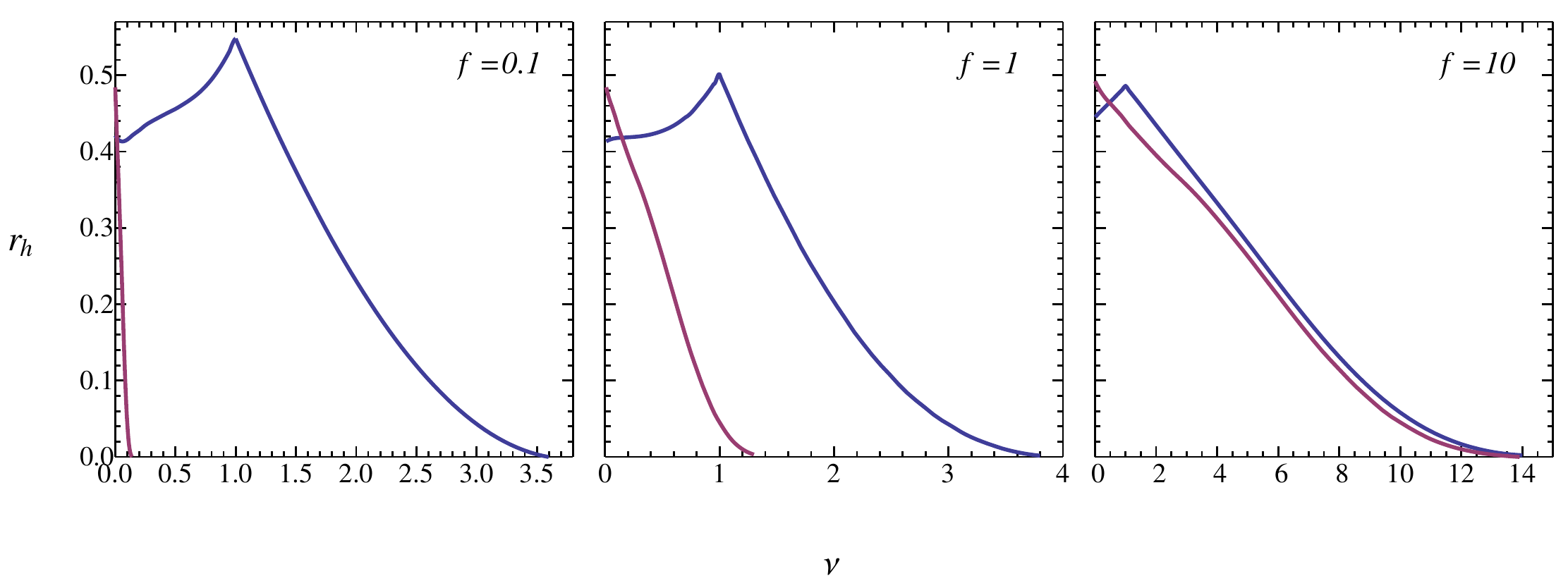}
\end{center}
\caption{\label{figure003}Chiral Symmetry Breaking: The wedge in the lower left below the red and blue lines  are the regions where the Abelian
 D5 brane and the D7 brane, respectively. have lower energies than the chiral symmetric D5 brane.  The horizontal axis is the filling fraction $\nu=
\frac{2\pi \rho}{NB}$ and the
vertical axis is the horizon radius (which is proportional to the temperature), in units of magnetic field.  The parameter $f=\frac{2\pi N_5}{\sqrt{\lambda}}$ is proportional to the number of D5-branes. Plots for three different values of $f$ are shown.   }
\end{figure}
If we keep the charge density and the temperature
at zero and introduce a constant external magnetic field, the D5 brane geometry changes drastically \cite{Filev:2009xp}.
Near the boundary of $AdS_5\times S^5$, the D5 brane is still $AdS_4\times S^2$.  However,  as it enters the bulk
 of $AdS_5$,
it pinches off and ends before it reaches the Poincare horizon of $AdS_5$, forming what is called a Minkowski embedding. It can pinch off smoothly
without creating a boundary when a cycle  shrinks to zero size.  It is the $S^2$ which shrinks and chiral symmetry is spontaneously broken.
Moreover, a  Minkowski embedding has a charge
gap due to the fact that charged excitations are open strings which must be suspended between the worldvolume and the Poincare horizon.  When the worldvolume does not reach the Poincare horizon,
these strings have a minimal length and therefore a gap in their energy spectrum.  
Thus, the strong coupling limit at $\nu=0$ has an incompressible, charge-gapped state.
This phenomenon is interpreted as the strong coupling manifestation of magnetic catalysis of chiral symmetry
breaking.  It is reasonable to conjecture that it is
precisely the continuation to strong coupling of the formation of a gap at $\nu=0$
in the neutral Landau level which we discussed at weak
coupling.  The symmetry breaking pattern is $SO(3)\times SU(N_5)\to SO(2)\times SU(N_5)$.
When a  charge density and a temperature are turned on the chiral symmetry breaking phase
survives for some range of these parameters, but it is eventually restored if the density or
temperature become large enough.\footnote{The fact that the spontaneous
breaking of a continuous symmetry in 2+1-dimensions survives at
finite temperature would seem to contradict the Coleman-Mermin-Wagner theorem in the field theory.  This is a typical artifact of the
large $N$   limit.  }   The chiral symmetric phase of the D5 brane which is found at large enough temperature or density is simply one where the
 worldsheet  is a product metric of the maximal $S^2$ embedded in $S^5$ and an asymptotically $AdS_4$ space,
together with a configuration of worldvolume electro-magnetic fields which are needed to create
the dual of the field theory with nonzero density and magnetic field.

\begin{figure}
\begin{center}
\includegraphics[scale=.65]{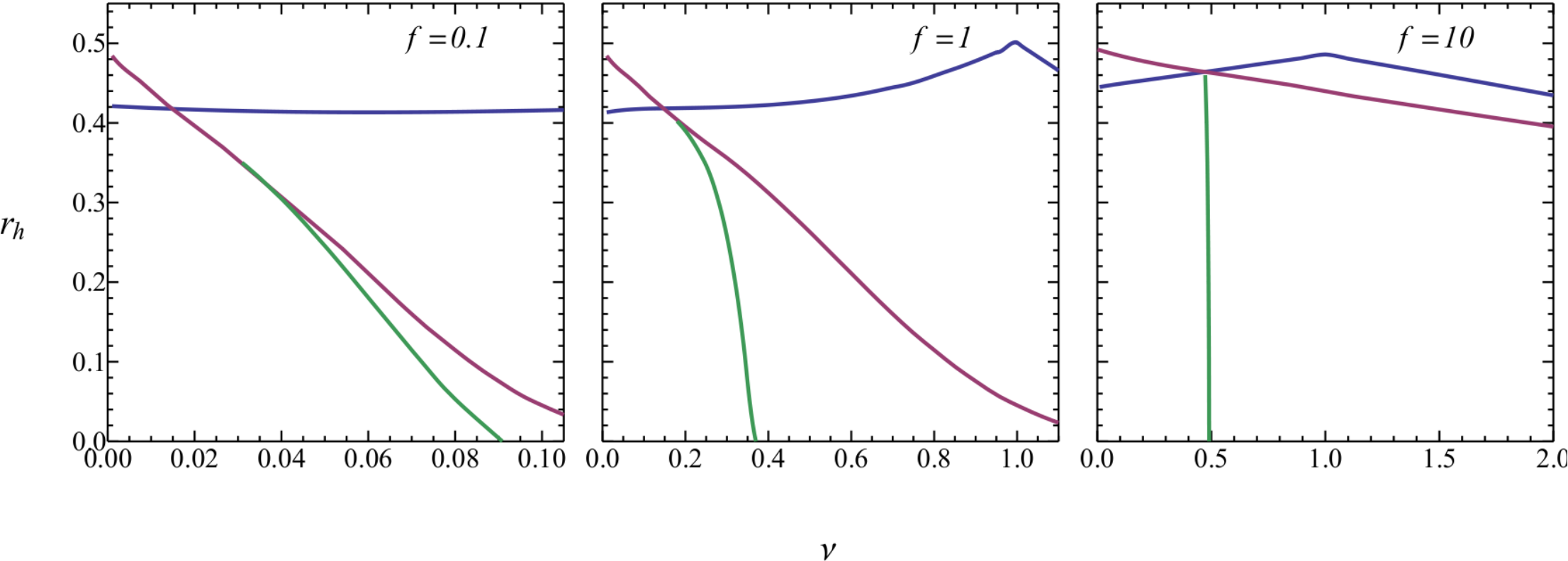}
\end{center}
\caption{\label{figure004}Phase diagram extracted from numerical data:
The red and blue lines are taken from figure \ref{figure003}. They are
lines where the chiral symmetric D5 brane has the same energy as the D5 brane (red) and the D7 brane (blue). The chiral
symmetric phase is always more stable to the right and toward the top of the figure.
The green line is where the Abelian D5 brane and the D7 brane have the same energy with the Abelian D5 preferred
to the left and the D7 preferred to the right. The axes and values of $f$ are as in  figure \ref{figure003}.}
\end{figure}

 The simple  chiral symmetry breaking solutions of the D3-D5 system have been studied extensively in a number of
contexts
\cite{Evans:2008nf}-\cite{Grignani:2012qz}.    Their distinguishing feature  can be characterized as ``Abelian'', in that
the dynamics of each D5-brane in the stack of D5 branes is treated independently and their
behaviors are all identical.  The non-Abelian nature of the worldvolume theory does not play
a role. The D5 branes remain coincident and have unbroken $SU(N_5)$ symmetry.
The phase diagram of these Abelian solutions is well known.  It is reproduced in the numerical
results of the present paper and corresponds to the red curves in figure  \ref{figure003}, \ref{figure004},  and
\ref{Compositeplot}.
More specifically, the lower-left-hand wedge in figure \ref{figure003} is the region where
the ``Abelian'' chiral symmetry breaking solutions of the D5-brane are stable.   At the red line, the chiral symmetric competitor
takes over, in the sense that it has lower energy.  (The same red curve re-appears in figure \ref{figure004} and
figure \ref{Compositeplot}.)
A more detailed discussion with more details about this phase transition
is given by Evans et.~al.~\cite{Evans:2010hi}.

The D7 brane is an alternative solution of the D5 brane theory.\footnote{An alternative solution of a D7 brane theory, called D7', has
been studied extensively \cite{Bergman:2010gm}-\cite{Jokela:2012vn}.   It also exhibits incompressible Hall states but with Hall conductivities that are given by
irrational numbers.   The main difference between that solution and the one that we study here is the behavior at the boundary of
$AdS_5$.   The D7 that we consider collapses to a D5 brane there, and should therefore be thought of as a solution of the D3-D5 system
whereas D7' remains a D7 brane.  }  It can be thought of as a ``non-Abelian''
configuration of D5 branes which is approximated by a D7 brane \cite{Kristjansen:2012ny}.
The stability region for the D7 brane in the temperature-density plane
is similar to that of the Abelian D5 brane, but somewhat larger.  It has less energy than the chiral symmetric
competitor (the same competitor
as for the Abelian D5 brane) to the lower left of the blue line in figure \ref{figure003}.
The reader should beware that figure \ref{figure003} does not compare the relative energies of the Abelian D5 and
the D7 branes.  This will be done in figure \ref{figure004}.

We now know of three competing solutions of the D3-D5 system, the Abelian D5 brane, the D7 brane and the chiral symmetric
solution. (As we shall see later, for $\nu>1$ we will in addition have the possibility of composite solutions.)
To decide which is the preferred one at a given value of the temperature and filling fraction, we must compare their free energies.
The free energy is a
function of the temperature, the charge density, the magnetic field and the number of D5-branes, $N_5$.
We will use a system of units where the magnetic field is equal to one.  This leaves a normalized temperature, the filling fraction $\nu$
and the parameter
\begin{align}\label{definef1}
f\equiv \frac{2\pi N_5}{\sqrt{\lambda}},
\end{align}
as the variables which define the thermodynamic problem.  We have done a numerical calculation of the energies of the three competing
solutions.  Before we begin to describe it, we warn the reader that, later on, we will find a completely
different structure that takes over when the filling fraction $\nu$ is greater than or equal to one,
and where the temperature is low. 
Thus, in the end, the following discussion will only apply to the region $0<\nu<1$.
We will ignore this fact for now and will pursue the following discussion of the relative
stabilities of the three solutions that we know about so far for all values of $\nu$.

In figure \ref{figure004}, the regions where the Abelian D5 brane and the D7 brane are more stable than the chiral symmetric
solution are displayed for three values of $f$.   In all cases, the chiral symmetric solution is stable in the upper right-hand part of the diagram.
The Abelian D5 brane is stable in the lower left, below the green line, that is, in all cases, for sufficiently small filling fraction and low temperature.
As we increase temperature or filling fraction to the green line, there is a phase transition and, beyond the green line,  the D7 brane becomes the energetically preferred solution.   It remains so until we approach the blue line  where the chiral symmetric phase becomes more stable and
chiral symmetry is restored.  This part of the blue line is beyond the edge of the figures in \ref{figure004} but can be seen in figure \ref{figure003}.
In summary, at low temperatures, as we increase the filling fraction from zero, generically there are three phases.  First, at low density is the Abelian D5 brane. At some value of the density, there is a phase transition to the D7 brane.  Then, at a larger density yet, there is a phase transition to the chiral
symmetric state.

At low temperatures, the phase transition at the blue line in figures \ref{figure003} and
\ref{figure004} is likely always of first order.   At zero temperature, the chiral symmetric phase
can be shown analytically to be meta-stable beyond a critical density,
which is also known analytically \cite{Jensen:2010ga}
\begin{align}\label{bkt}
\nu_{\rm crit}\approx 1.68 f.
\end{align}
In all of the cases where we have computed it, the blue line always occurs at higher values of $\nu$.\footnote{For very large values of $f$, we have
observed that the red and
blue lines come closer together and it is possible that, for sufficiently large $f$, they will coincide. In that case, there could be a large $f$ region
where the the BKT-like transition from either the abelian D5 or D7 brane to the chiral symmetric phase would still exist.}  This means that there is always a region where
the D7 brane has lower energy than a meta-stable chiral symmetric phase.  In this region, there must be an energy barrier between the phases.
This indicates a first order phase transition.
In the case of the Abelian solution at zero temperature, it is known that the transition to the chiral symmetric phase occurs precisely at $\nu_{\rm crit}$
and it is BKT like \cite{Jensen:2010ga}.  This is the intersection of the red line with the horizontal axis in figures \ref{figure003} and \ref{figure004}.
It is a beautiful and rare example of a non-mean-field phase transition for probe D branes.
However, in all of the cases that we have studied, including those in figure \ref{figure004},
the green line occurs at values $\nu$ smaller than  $\nu_{\rm crit}$ and the D7 brane is more stable than either the Abelian D5 or the chiral symmetric D5 in this region - there is no BKT-like transition in these cases.

The three phases that we have discussed so far have distinct symmetry breaking patterns.  The Abelian D5 brane phase breaks the $SO(3)$  symmetry to $SO(2)$, but preserves the $SU(N_5)$ flavor symmetry.    The D7 brane phase breaks both  the $SO(3)$  symmetry and the   $SU(N_5)$ leaving only a subgroup
of simultaneous transformations in $SO(3)$ and an $SO(3)$ subgroup of $SU(N_5)$.  Then the chiral symmetric phase retains both the $SO(3)$  and the $SU(N_5)$.    This pattern of symmetry breaking  is
one of the predictions of the holographic model.   However, as we have warned the reader. further developments that we shall outline below will cut
this scenario off when the filling fraction reaches $\nu=1$.  It is only the behavior which occurs in the interval where $0<\nu<1$ which will turn out to be a prediction of what we have done so far.\footnote{Of course, in a real
two dimensional electron gas, the physics of some of 
the region that we are talking about here is dominated by other effects such as impurities and localization.  Our prediction could still appy to the symmetries of the Hall plateaus which
are formed by these other effects.  The other possibility is the clean limit which is thus far proven
difficult to achieve, even in graphene.  }
As we can see by inspecting figures \ref{figure003} and \ref{figure004},
the first transition from Abelian D5 to D7 typically occurs in this interval whereas the second, from D7 to chiral symmetric D5 does not.

The Abelian D5 brane has the feature that, once a non-zero charge density is turned on, it can no longer have a Minkowski embedding.  This means that
the theory no longer has a charge gap. Without a magnetic field, this would be natural.
The analog at weak coupling is a finite density of fermions which create a Fermi surface.  There are always
low energy excitations of a Fermi surface.  Such a system is not gapped, and this is also what is seen for the Abelian
D5 brane.  However,  the Abelian D5 brane
also remains ungapped in an external magnetic field, for any value of the magnetic field and any nonzero density.  In other words, 
besides the $\nu=0$ state, the Abelian D5 brane solutions contain no incompressible states at non-zero filling fractions, even in arbitrarily
strong magnetic fields.

The D7 brane, on the other hand, can have incompressible states at  special nonzero values of
the charge density \cite{Kristjansen:2012ny}.     The D7 brane should properly be regarded as a non-Abelian configuration of D5-branes.  It arises in the D5 brane theory when the transverse coordinates
of the $N_5$ embedded D5 branes, which are $N_5\times N_5$ matrices,  form a fuzzy sphere.\footnote{  $S^5$ is the locus of $n_1^2+\ldots+n_6^2=1$ for real numbers
$(n_1,...,n_6)$.  The two $S^2$'s are loci of  $n_1^2+n_2^2+n_3^2=\sin^2\psi$ and $n_4^2+n_5^2+n_6^2=\cos^2\psi$.
All D5 and D7 brane solutions which we discuss wrap the  first sphere,   $(n_1,n_2,n_3)$, and its $SO(3)$ isometry is preserved.
For the D5 brane, $(n_1,n_2,n_3)$ are longitudinal coordinates and $(n_4,n_5,n_6)$ are transverse coordinates.
The chiral symmetric
D5 brane sits at a higher symmetry point $n_4=n_5=n_6=0$ (or $\psi=\tfrac{\pi}{2}$)
and preserves the second $SO(3)$ symmetry.   The Abelian D5 solution
has some of the $(n_4,n_5,n_6)$ non-zero.  This
breaks the second $SO(3)$ (and preserves $SU(N_5)$).
The D7 brane is a non-Abelian D5 brane where $(n_4,n_5,n_6)$ are $N_5\times N_5$ matrices  in the $SU(N_5)$ Lie algebra
and which form an
irreducible representation of an $SO(3)$ subalgebra of $SU(N_5)$, \begin{align}\label{so3algebra}[n_a,n_b]=i\epsilon_{abc}n_c~.\end{align}
The D7 brane  preserves an $SO(3)$ which is a combination of the $SO(3)$ of the second $S^2$ and
the $SO(3)$ subgroup of $SU(N_5)$.  A time reversal invariant fermion mass operator  (and order parameter for chiral symmetry breaking)
which is invariant under the residual symmetry
would be $m\bar\psi n_a\sigma^a\psi$.  In the classical description as a D7 brane wrapping the second $S^2$ with $N_5$
units of monopole flux, the unbroken symmetry is the magnetic translation group on the $S^2$.}
 It would be interesting to understand  this ``non-Abelian'' configuration of the  D5 brane better from this point of view.
We shall not do this in the following.
Instead, we simply approximate it by a classical D7 brane which wraps both of the $S^2$'s in $S^5$.  
The second $S^2$ is the classical limit of the fuzzy sphere.  The D7 brane remembers its origin as
$N_5$ D5 branes by supporting a magnetic flux with Dirac monopole number $N_5$ on the second $S^2$.
 This approximation should be good when the number of D5 branes, $N_5$, is much greater than one,
but still much less than $N$.
In the asymptotic region near the boundary of $AdS_5\times S^5$ the D7 brane has geometry $AdS_4\times S^2\times S^2$
where one $S^2$, the same one which is wrapped by the Abelian D5 brane,
is nearly maximal and the other $S^2$ shrinks to zero size as the boundary is approached.  This sphere has magnetic flux and
when it shrinks to zero size it leaves a singular magnetic source.  This can be regarded as a point where a D5 brane is attached
to the D7 brane and it occurs precisely at the boundary of $AdS_5$.

In figures \ref{figure003} and \ref{figure004} there are peaks of the blue curves at $\nu=1$.  This is the special state of the D7 brane, where it has
a Minkowski embedding and is incompressible with a charge gap.  At low enough temperature, this state is energetically preferred over
its competitors in the entire range of the parameter $f$ that we have been able to study.
Its existence also allows us to find an incompressible state for higher integer values of
$\nu$.  This is gotten by simply taking $\nu$ D7 branes, each with filling fraction equal to one.  As well as the charge density,
which they share equally, the D7 branes must share the $N_5$ D5 branes between them.  A numerical computation
in reference \cite{Kristjansen:2012ny} indicated that, at least for $\nu=2$, first of all, the energy is minimized when the D5 branes  are
shared equally and secondly, the energy of two gapped $\nu=1$ D7 branes with $f$ shared
equally between them   
is less than the energy of one ungapped
$\nu=2$ D7 brane or Abelian D5 brane.
The second of these results tells us that the $\nu=2$ gapped state with 2 D7 branes is preferred over
the other possible ungapped states.  The first one tells us that the gapped state is a state with two coincident gapped D7 branes. 
We conjecture but have not checked that this pattern persists to higher values of $\nu$.  Here, we shall assume
that when the charge density is shared equally, the branes also prefer to share $N_5$ D5 branes equally.  Indeed, this state has
more symmetry than the alternatives, since the $\nu$ D7 branes are identical and coincident, and therefore have an unbroken internal
gauge symmetry $SU(\nu)$. This would be an unbroken global symmetry of the field theory dual. Of course, in the strict sense,
it can only happen
if $\nu$ is an integer divisor of $N_5$.   However, in the large $N_5$ limit that we are considering,
the $N_5$ D5 branes can always be split equally to precision $\frac{1}{N_5}$ and the residual symmetry
would be there to a very good approximation.
This symmetry would be  a subgroup $SU(\nu)\subset SU(N_5)$ which (in addition to some
$ SO(3)$'s), survives dynamical symmetry breaking
by D7 branes.
Its existence can be regarded as a prediction of our hypothesis for finding the charge gapped state
with integer filling fraction $\nu$.

Now, we are ready to take the next step and understand the ungapped states in the region between integer filling fractions, say the region
$1\leq \nu\leq 2$.  At $\nu=1$ the stable state is the gapped D7 brane.  If we increase $\nu$ slightly, we might expect that the lowest
energy state is a composite brane made from the same gapped D7 brane
and either an ungapped Abelian D5 brane or an ungapped
D7 brane where, in both cases,  the second, ungapped brane takes on a share of the filling fraction, 
$\nu-1$. In addition to this, the gapped and ungapped branes must share the $N_5$ D5 branes between them.
Exactly how $N_5$ is distributed between the branes in the composite system
is a dynamical question which we shall solve numerically in a few cases.\footnote{In the non-Abelian picture, the
transverse matrix-valued coordinates of the gapped D7-ungapped D7 brane composite would have the $n_a$  in
equation (\ref{so3algebra}) block-diagonal,
$$
n_a=\left[ \begin{matrix} L_a^{(1)} & 0 \cr 0 & L_a^{(2)}\cr \end{matrix} \right]
$$
where $L_a^{(1)}$ is a $n\times n$ and $L_a^{(2)}$ is an $(N_5-n)\times(N_5-n)$ irreducible representation of $SO(3)$,
respectively. For the gapped D7-ungapped Abelian D5 brane composite, $L_a^{(2)}$ is replaced by $0$.}

Our investigation shows that, which ungapped brane is stable depends on the total $N_5$ through the parameter $f$ defined in
(\ref{definef1}).  If $f$ is big
enough, the Abelian D5 brane wins, and the state just above  $\nu= 1$  is a hybrid of the gapped D7 brane and the ungapped Abelian D5 brane.
Then, there is a
phase transition in this intermediate region $\nu\in[1,2]$, at a critical value of $\nu$, to a state which is a composite of the gapped D7 and an ungapped D7 brane.
As $\nu$ is increased further, and $\nu=2$ approached from smaller values of $\nu$, the state should be the gapped D7-ungapped D7
brane composite. When $\nu=2$ is reached, as we have discussed above, it becomes two coincident gapped D7 branes, each with $\nu=1$.
At smaller values of $f$, our results indicate that the state just above $\nu=1$ is immediately a composite of the gapped D7 brane
and the ungapped  D7  brane. The Abelian D5 brane does not appear at all in the interval $1\leq\nu\leq2$.

A similar pattern of  composite branes is repeated in the intervals between larger integer values of $\nu$.  We have 
investigated this 
by numerical computation and have found that it is indeed the case.  We have
explicit numerical solutions 
up to the interval $8\leq\nu\leq 9$.  We currently have no evidence that the pattern
stops.  We also find that, even when $f$
is large enough that the Abelian D5 brane phase exists just above the lower integer $\nu$'s, this is
so only of the smaller values of $\nu$.  At higher
integers, the Abelian D5 phase ceases to occur and integer $\nu$ D7 branes immediately become
a composite of the gapped D7 branes and an ungapped D7 brane when $\nu$ is increased
beyond the integer value. 

In summary, in the defect quantum field theory that we are studying, 
when the magnetic field is turned on, for any value of the field strength, the chiral symmetry is broken in that there is always a chiral condensate.
However, there is a charge gap
only when $\nu$ is an integer, either positive or negative, and for values of $\nu$ with magnitude no bigger than $N_5$.  The series could truncate before it gets
to $\nu=\pm N_5$. We have not seen numerical evidence for this truncation, mainly due to the fact that our analysis considers
very large $N_5$ and smaller values of $\nu$.
 Between the integer values of $\nu$, even though there is a nonzero chiral condensate, there is no charge gap.

The symmetry breaking patterns are then quite interesting.  Let us begin with the case where $f$ is small enough that the only
composite branes are gapped D7-ungapped D7 branes. Let us begin at $\nu=0$.
There, at $\nu=0$, as soon as the magnetic field is turned on,
the SO(3)  is spontaneously broken to an $SO(2)$ subgroup.  
The full $SU(N_5)$ symmetry is preserved there.
As $\nu$ is increased, this symmetry breaking pattern 
persists up to a phase transition at a critical value of  $\nu$
between zero and one, where there is a phase transition.
At that phase transition, the system goes to a phase where the only symmetry which survives
is in a
diagonal subgroup of $SO(3)$ and an $SO(3)$ subgroup of $SU(N_5)$.     This symmetry breaking pattern then
persists until we reach $\nu=1$, where it is also the symmetry of the charge gapped state which appears at $\nu=1$.
Then, when we increase $\nu$ beyond one, the pattern changes again.  The composite gapped D7-ungapped D7 brane that is stable there
preserves  two diagonal
$SO(3)$'s, one for each D7 brane.  These consist of $SO(3)$ rotations combined with rotations in two commuting $SO(3)$ subgroups
of $SU(N_5)$.   When we reach $\nu=2$, this symmetry is enhanced once again.  The diagonal $SO(3)$'s become
degenerate and they are transformed into each other by an additional $SU(2)$ subgroup of $SU(N_5)$.
That is, at $\n=2$, out of the original $SO(3)\times SU(N_5)$,
the symmetry
which survives  is  $[SO(3)]^2\times SU(2)$.
When we increase
$\nu$ to values just above two, the stable solution being a two gapped D7-one ungapped D7 composite,
the symmetries that existed at filling fraction two are still there and, in addition,
another new diagonal $SO(3)$ emerges so that the total is $[SO(3)]^2\times SO(3)\times SU(2)$.
The latter is the symmetry of the third, gapped D7 brane.
When we reach $\nu=3$, the third diagonal $SO(3)$ becomes degenerate with the first two, so that
$[SO(3)]^2\times SO(3)\times SU(2)$ becomes $[SO(3)]^3\times SU(3)$.  As far as we have investigated,
this pattern repeats
itself as we proceed to higher values of $\nu$.   At $\nu=n$, out of the original $SO(3)\times SU(N_5)$,
the symmetry
which survives  is  $[SO(3)]^n\times SU(n)$. When $\nu$ is just above $n$, this gets an additional $SO(3)$ to become
$[SO(3)]^n\times SO(3)\times SU(n)$.  When we reach $\nu=n+1$ this is enhanced again to $[SO(3)]^{n+1}\times SU(n+1)$.
Of course, we know this reliably
only when $n<<N_5$.  To study   what happens
for larger values of $n$ is beyond our current ability, but would indeed be very interesting.

If the parameter $f$ is larger, the additional composite phase, where there are $n$ gapped D7 branes and an Abelian D5 brane,
inserts itself in some of the regions just above $\nu=n$.   We find that this is so, at least for big enough values of $f$ and
for small enough values
of $n$. We have seen that, for $f=10$, this happens for $\nu=1,2,3$ and it ends at $\nu=4$.  Thereafter the states are always composites of
D7 branes.   This phase involving D5 branes breaks the  $SO(3)$ of the second $S^2$ to $SO(2)$, and it leaves its share of the $SU(N_5)$ symmetry intact.
It therefore has residual symmetry $[SO(2)]^n\times SU(n)\times SU(N_5-N_5^0)$
where $N_5^0$ is the number of D5 branes that are absorbed by the $n$ gapped D7 branes in the composite, leaving a
stack of $N_5-N_5^0$ D5-branes for the Abelian D5 brane part of the composite.
Then, somewhere in the interval $\nu=[n,n+1]$ we expect a phase transition where the ungapped Abelian D5 brane blows up to
an ungapped D7 brane, so that $[SO(2)]^n\times SU(n)\times SU(N_5-N_5^0)
\to [SO(3)]^n\times SO(3)\times SU(n)$. At some big enough value of $n$, the intermediate composites with Abelian D5 branes cease
to exist and the pattern of the preceding paragraph takes over.

These symmetry breaking patterns can be considered a prediction of our holographic model and it is interesting to ask whether
they are relevant to any realistic system.  Aside from the supersymmetric system that is modelled directly, there is some hope that
the model also captures some of the physics of any electronic system with degenerate Landau levels and a strong repulsive
interaction.  If interactions are ignored entirely,  $2N_5$ degenerate Landau levels have an effective $SU(2N_5)$ symmetry.
In our model, the interactions on the other hand, have only a smaller symmetry,
$SO(3)\times SU(N_5)$.  (We are also ignoring the other, first $SO(3)$,  which transforms the first $S^2$ in the string theory and
acts on the scalar fields in the weakly coupled field theory.)   We are only able to analyze the situation where $N_5$ is large.  We might
wonder whether some aspect of the symmetry breaking pattern survives for small values of  $N_5$.
Then, the generic prediction is that the $\nu=0$ state, that is
the one where half of the states are filled,  has distinctly different symmetry from all of the others.

The most realistic possibility is $N_5=2$ which could match the symmetries (spin times valley) of graphene or the
a bilayer quantum Hall system where the SO(3) is spin symmetry and the SU(2) transforms the layer index.  In both of these cases,
the valley/layer symmetry is only approximate.  In graphene, the long ranged Coulomb interaction is $SU(4)$ symmetric and short
ranged parts reduce this symmetry to $SO(3)\times Z_2$, which is sometimes approximated by  $SO(3)\times SU(2)$  where further, weaker
interactions break the $SU(2)$ \cite{quantuhallferromagnet3.5}.

We could ask how our symmetry breaking patterns would be seen in  the weak coupling states 
where the charge neutral Landau level is fractionally filled.    
Let us go back to weak coupling for a moment. 
  Denote the completely empty four-fold degenerate Landau level as $|0>$  and
the electron creation operator as $\psi^\dagger_{Pab}$ where $P$ denotes a state in some basis of the degenerate single-electron states of the
 Landau level and $a,b$, each taking values $\uparrow,\downarrow$
are valley/layer  and spin indices.   To get a translation invariant state, we must create an electron within each degenerate state of the
Landau level, that is we must fill all of the states denoted by $P$.
 We begin by half-filling the Landau level to get the
$\nu=0$ state.   That state has  $SO(3)$ symmetry broken to $SO(2)$, but still has good $SU(2)$   symmetry.  This should be the
symmetry pattern of the $\nu=0$ plateau.  The weak coupling state that does this is
$$
\prod_{P}\psi^\dagger_{P\uparrow\uparrow}\psi_{P\uparrow\downarrow}^\dagger~|0>
$$
which is, for each state $P$ in the Landau level, a valley triplet and spin singlet.  It thus breaks the
valley symmetry and preserves the spin symmetry.  (The inverse is also possible, where the spin symmetry
is broken and the valley symmetry survives. Here we are
not  distinguishing spin and valley symmetries.)   

Then, consider the one-quarter and three-quarter filled states.  For graphene, these are the $\nu=- 1$ and $\nu=1$ states, respectively.
Quarter filling is achieved by creating an
electron in one quarter of the zero mode states.  A simple
candidate for such a state is
$$
\prod_{P}\psi^\dagger_{P\uparrow\uparrow}~|0>
$$
which is both spin and valley polarized.   
\footnote{If graphene, say, were fully $SU(4)$ symmetric, the $SU(4)$ could be used to rotate this
state to any other choice, so if the dynamics were $SU(4)$
invariant, this would be the most general state.  
However, if the  interactions are not fully $SU(4)$ invariant, but are symmetric under $SO(3)\times SU(2)$ instead, this state would break both
the $SO(3)$ and $SU(2)$ symmetries.  What is more, it is not the unique choice for
a quarter filled state.  The holographic states suggest the alternative (\ref{state}).}
This state breaks both the spin and valley symmetry.   This is a different symmetry breaking
pattern than we found for our holographic state with $\nu=1$.  We can make a many body state
with a symmetry breaking pattern which matches the  holographic state.  
   It would be the state
\begin{align}\label{state}
\prod_{P}\frac{1}{\sqrt{2}}\epsilon^{ab} \psi^{\dagger}_{Pab}~|0>=
\prod_{P}\frac{1}{\sqrt{2}}(\psi^\dagger_{P\uparrow\downarrow}-\psi^{\dagger}_{P\downarrow\uparrow})~|0>
\end{align}
This state is neither spin nor valley  polarized.   It is a singlet under a simultaneous spin and valley rotation, and a triplet under
a spin rotation and a simultaneous valley  inverse rotation.    Since, for the fermion zero mode Landau level that we are discussing, 
the wave-function of the zero modes in a specific valley also occupy only one of the graphene sublattices \cite{Semenoff:1988ep}, 
the other valley occupying
the other sublattice, a flip of the valley index corresponds to a translation which interchanges the sublattices.  Since the state (\ref{state}) is
left unchanged by a simultaneous flip of valley and spin indices, this state is then an anti-ferromagnet. 

There is a state at three-quarters filling that has similar symmetries,
$$
\prod_{P}\frac{1}{\sqrt{8}}\epsilon^{fa}\psi^\dagger_{Pab}\epsilon^{bc}\psi^\dagger_{Pcd}\epsilon^{de}
\psi^\dagger_{Pef} ~|0>
$$
This state is also spin and valley unpolarized. The fact that the integer $\nu\neq 0$ states are that way
is a generic feature of the holographic model.  Here, we see that it suggests a particular state
for both $\nu=-1$ and $\nu=1$. It would be interesting to see if this suggestion is realized in graphene or multilayer Hall systems.
Recent experimental results seem consistent with this picture \cite{kim}.

The remainder of the paper discusses the details of the work that has been summarized in this introduction.   In section 2 we discuss
the mathematical problem of embedding D5 and D7 branes in $AdS_5\times S^5$.   In section 3, we discuss the boundary conditions and the
asymptotic behavior of the solutions that we are looking for. In section 4 we discuss the details of both the technique and results
of our numerical computations.   In section 5 we conclude and we discuss directions for further work.

\section{The geometric set-up}

We shall study D5 and D7 probe branes at finite temperature and density.   We will embed them in the asymptotically $AdS_5\times S^5$
black hole background using coordinates where the metric has the form
\begin{align}
ds^2= \sqrt{\lambda}\alpha'&\left[ r^2 (-h(r)dt^2+dx^2+dy^2+dz^2) + \frac{dr^2}{h(r)r^2}+\right. \nonumber \\
&\left. +d\psi^2+\sin^2\psi(d\theta^2+\sin^2\theta d\phi^2)+ \cos^2\psi (d\tilde\theta^2+\sin^2\tilde\theta d\tilde\phi^2)\right],
\label{ads5metric}\end{align}
Here,  the coordinates of $S^5$ are a fibration of the 5-sphere by two
2-spheres over the interval $\psi\in[0,\pi/2]$. Furthermore,
$(t,x,y,z,r)$ are coordinates of the Poincare patch of $AdS_5$ and
$$
h(r)=1-\frac{r_h^4}{r^4}
~,
$$
with $r_h$  the radius of the event horizon.  The Hawking temperature is $T=  r_h/\pi$.
The Ramond-Ramond 4-form of the IIB supergravity background takes the form
\begin{align}\label{4form}
C^{(4)}=  \lambda{\alpha'}^2\left[h(r)
r^4dt\wedge dx\wedge dy\wedge dz+ \frac{c(\psi)}{2}d\cos\theta\wedge d\phi
\wedge d\cos\tilde\theta\wedge d\tilde\phi\right],
\end{align}
Here,
$
\partial_\psi c(\psi)=8\sin^2\psi\cos^2\psi
$
and for later convenience we choose
\begin{align}
c(\psi)= \psi - \frac{1}{4}\sin4\psi -\frac{\pi}{2}.
\label{cpsi}
\end{align}
The choice of integration constant is a string theory gauge choice and our
results will not depend on it.

We will study D5 and D7 branes as well as some composite systems
thereof in this background using the probe approximation where the number
of probe branes $N_5$ and $N_7$ is much smaller than the number, N, of
D3-branes.
The world volume coordinates of our probe branes will be as
given in the table below.
\begin{align}
\\
\nonumber
\boxed{\begin{array}{rcccccccccccl}
  & & t & x & y & z & r & \psi & \theta & \phi & \tilde{\theta} & \tilde{\phi} &\\
& D3 & \times & \times & \times & \times & & &  & & & & \\
& D5 & \times & \times & \times &  & \times  &  & \times & \times &  & &  \\
& D7 & \times & \times & \times &  & \times &  & \times & \times & \times &\times &   \\
\end{array}}
\nonumber
\\
\\
{\rm\bf Table~1:~D3,~ D5~ and~ D7~world~volume~coordinates}
\nonumber
\end{align}

\subsection{Probe D5 branes}

Probe D5 branes are described by the DBI plus WZ actions, i.e.
\begin{align} \label{dbi5}
S_5=\frac{ T_5}{g_s} N_5 \int d^{\:6}\sigma
\left[- \sqrt{-\det( g+2\pi\alpha'{\mathcal F}_5)} +2\pi\alpha'
C^{(4)}\wedge {\mathcal F}_5\right],
\end{align}
where $g_s$ is the closed string coupling constant, which is related to the ${\mathcal N}=4$
Yang-Mills coupling
by $4\pi g_s =g_{YM}^2$, $\sigma^a$ are the coordinates of the D5 brane
worldvolume, $g_{ab}(\sigma)$ is the induced metric of the D5 brane,
${\mathcal F}_5$ is the worldvolume gauge field and
\begin{align}\label{t5}
T_5=\frac{1}{(2\pi)^5{\alpha'}^3},
\end{align}
is the D5 brane tension.
The Wess-Zumino action will not contribute
to the D5 brane equations of motion for the types of embeddings that we
will discuss here. The overall factor of $N_5$ denotes the number of
D5 branes. We are here assuming that the non-Abelian $U(N_5)$ gauge
symmetry structure of multiple $N_5$ branes plays no role. We shall take the
world volume gauge field strength to be of the form
\begin{align}
2\pi\alpha'{\mathcal F}_5=\sqrt{\lambda}\alpha'\left[ \frac{d}{dr}a(r)dr\wedge dt + bdx\wedge dy \right].
\end{align}
Hence, we have a constant external magnetic field
\begin{align}\label{externalmagneticfield}
B= \frac{\sqrt{\lambda}}{2\pi}~b,
\end{align}
and a charge density $\rho$
 $$
\rho =
\frac{1}{V_{2+1}}\frac{2\pi}{\sqrt{\lambda}}
\frac{\delta S_5}{\delta \frac{d}{dr}a(r)}\label{chargedensity}.
$$
It is well known that there exists an embedding of the
D5 brane with the world volume coordinates
$(t,x,y,r,\theta,\phi)$ for which  $\tilde{\theta}$,
$\tilde{\phi}$ and $z$ are constant and for which  $\psi=\psi(r)$
depends only on $r$. For this embedding the world volume
metric can be written
as
\begin{align}\label{D5metric}
ds^2 = \sqrt{\lambda}\alpha'\left[ r^2 (-h(r)dt^2+dx^2+dy^2)
+\sin^2\psi(d \theta^2+\sin^2 \theta d\phi^2)\right.. \nonumber \\
+\left.\frac{dr^2}{h(r)r^2}\left(  1+h(r)\left( r\frac{d\psi}{dr} \right)^2 \right)
\right],
\end{align}
and the DBI action becomes
\begin{align}\label{ansatzaction5}
S_5=-{\mathcal N}_5N_5\int_0^\infty  dr~  2\sin^2\psi\sqrt{b^2+r^4}\sqrt{1+h(r)\left(r\frac{d\psi}{dr}\right)^2-\left(\frac{da}{dr}\right)^2},
\end{align}
  where, using (\ref{t5}),
\begin{align}\label{N5}
{\mathcal N}_5=\frac{T_5}{g_s} (\sqrt{\lambda}\alpha')^3(2\pi)V_{2+1}=\frac{2\sqrt{\lambda}  N}{(2\pi)^3}V_{2+1}.
\end{align}
The factor
$ (\sqrt{\lambda}\alpha')^3$  comes from the overall factor in the worldvolume metric
in equation (\ref{D5metric}), the factor of $(2\pi)$ is from the integral over the worldvolume two-sphere.\footnote{It is  half of the volume of the 2-sphere.  The other factor of 2 is still in the
action in front of $\sin^2\psi$. This notation is designed to match with the D7 brane, which we shall study
in the next section, and to coincide with notation in reference \cite{Kristjansen:2012ny}.}  and
the integral over $(x,y,t)$ produces the volume factor $V_{2+1}$.
To finalize the description of the embedding of the D5-brane we should
determine the functions $\psi(r)$ and $a(r)$ by varying the action above.
In the process of variation one can use the boundary condition
\begin{align}
\lim_{r\to\infty}\psi(r)=\frac{\pi}{2},
\label{bc}
\end{align}
which, as we shall see later, is compatible with the equation of motion
for $\psi$.
Since the variable $a(r)$ enters the Lagrangian only via its derivative,
$a(r)$ is cyclic and can be eliminated in favor of an integration constant
using its equations of motion. The corresponding integration constant is
(up to another constant factor) equal to the
charge density $\rho$, hence
$$
\rho = const.
$$
Eliminating $a(r)$ via a Legendre transform, following the steps of
reference~\cite{Kristjansen:2012ny},
gives us the Routhian,
\begin{align}
{\mathcal R}_5=-\frac{{\mathcal N}_5N_5}{f}\left(\frac{2\pi B}{\sqrt{\lambda}} \right)^{\frac{3}{2}}\int_0^\infty d r  ~{\cal L}_5,
\end{align}
where
\begin{align}
{\cal L}_5=
\sqrt{4\sin^4\psi f^2(1+r^4)+ (\pi\nu)^2}\sqrt{1+h(r)\left(r\frac{d\psi}{dr}\right)^2}.
\label{Routhian5}
\end{align}
Here $r$ is a dimensionless variable, obtained by
rescaling $r\rightarrow r \sqrt{b}$, the quantity $f$ is related to the
total number of D5 branes, i.e
\begin{align}
f=\frac{2\pi}{\sqrt{\lambda} } N_5,
\label{definef}
\end{align}
and finally $\nu$ is the filling fraction
\begin{align}\label{definenu}
\nu = \frac{2\pi}{N}\frac{\rho}{B}.
\end{align}
To determine $\psi(r)$  we should now finally extremize the
Routhian keeping $\nu$ fixed. This leads to the following equation of
motion for $\psi(r)$
\begin{align}
\frac{h\left(r\frac{d}{dr}\right)^2\psi}{1+h\left({r\frac{d}{dr}\psi}\right)^2}+hr\frac{d}{dr}\psi\left[ 1+\frac{8r^4\sin^4\psi  f^2 }{4\sin^4\psi
  f^2(1+r^4)+(\pi\nu)^2}\right]  \nonumber \\
-2\frac{r_h^4}{r^4} r\frac{d}{dr}\psi\left[1+\frac{1}{1+h\left({r\frac{d}{dr}\psi}\right)^2}\right]
-\frac{ 8\sin^3\psi\cos\psi f^2 (1+r^4) }{4\sin^4\psi f^2(1+r^4) +
(\pi \nu)^2}=0.
\label{equationforpsi5}\end{align}
where, now $r_h$ is in magnetic units, i.e.\ it has been rescaled $r_h\to r_h/b^\frac{1}{2}$ so that
\begin{align}
r_h^2 = \pi^2 T^2\frac{\sqrt{\lambda}}{2\pi B}.
\label{hawkingtemperature}
\end{align}
Note that the D5-brane solutions will depend only on the ratio $\frac{\nu}{f}$ and on the temperature $T$ in magnetic units.

\subsection{Probe D7 branes}
For probe D7 branes the DBI plus WZ action reads
\begin{align}
S=\frac{ T_7}{g_s} \int d^{\:8}\sigma\left[- \sqrt{-\det(   g+2\pi\alpha' {\mathcal F}_7)} +\frac{(2\pi\alpha')^2}{2}
C^{(4)}\wedge  {\mathcal F}_7\wedge  {\mathcal F}_7\right],
\end{align}
where
\begin{align}\label{t7}
T_7=\frac{1}{(2\pi)^7{\alpha'}^4},
\end{align}
is the D7 brane tension. Notice that here we are considering a single
D7 brane. The world volume gauge field strength we take to be of the form
\begin{align}
2\pi\alpha'{\mathcal F}_7=\sqrt{\lambda}\alpha'
\left( \frac{d}{dr}a(r)dr\wedge dt + bdx\wedge dy +
\frac{f}{2}d\cos\tilde\theta\wedge d\tilde\phi\right).
\end{align}
The flux parameter, $f$,  is the parameter defined in equation (\ref{definef}).
It corresponds to $N_5$ Dirac monopoles on $\tilde S^2$. The magnetic field
and the charge density are again given by expressions~(\ref{externalmagneticfield}) and~(\ref{chargedensity}).
We will now be interested in the embedding of the D7-brane with world volume
coordinates $(t,x,y,r,\theta,\phi,\tilde{\theta},\tilde{\phi})$ and we know
that there exists an embedding for which $z$ is a constant and $\psi=\psi(r)$
is a function of $\psi$ only. The corresponding D7 brane world volume
metric reads
\begin{align}\label{D7metric}
ds^2= \sqrt{\lambda}\alpha'\left[ r^2 (-h(r)dt^2+dx^2+dy^2)+\frac{dr^2}{h(r)r^2}\left(1+h(r)\left(r\frac{d\psi}{dr}\right)^2\right)+\right. \nonumber \\ \left. +\sin^2\psi(d\theta^2+\sin^2\theta d\phi^2) +\cos^2\psi(d\tilde\theta^2+\sin^2\tilde\theta d\tilde\phi^2)\right],
\end{align}
and the action becomes
\begin{align}\label{ansatzaction}
S_7=-{\mathcal N}_7\int_0^\infty  dr\left[2\sin^2\psi\sqrt{(f^2+4\cos^4\psi)(b^2+r^4)}
\sqrt{1+h(r)\left(r\frac{d\psi}{dr}\right)^2-\left(\frac{da}{dr}\right)^2} \right.
\nonumber \\ \left. +  2\,
\frac{da}{dr} b c(\psi)\right],
\end{align}
  where, using (\ref{t7}),
\begin{align}\label{N}
{\mathcal N}_7= \frac{2\lambda  N}{(2\pi)^4}V_{2+1}.
\end{align}
Again to finalize the embedding we have to determine the functions $\psi(r)$
and $a(r)$ by varying the action. We will use the same boundary condition as
before, i.e.\ the one given in (\ref{bc}) which again will indeed be compatible
with the equations of motion for $\psi(r)$. In this connection it is
convenient that we have chosen
$c(\psi)$ as in equation~(\ref{cpsi}).
Similarly to before $a(r)$ is
a cyclic variable which can be eliminated using its equation of motion and the
corresponding integration constant is again equal to the charge density up
to a constant factor (different from the one of the D5 case). We will
be interested in the situation where we fix the
integration constants so that the charge density, $\rho$, is the same
for  D5 branes and D7 branes. After eliminating $a(r)$ via a Legendre
transformation as before we find the following Routhian
\begin{align}\label{Routhian}
{\mathcal R}_7&=-{\mathcal N_7}\left(\frac{2\pi B}{\sqrt{\lambda}}\right)^{\frac{3}{2}}\int_0^\infty d r  {\cal L}_7,
\end{align}
with
\begin{align}
{\cal L}_7=\sqrt{4\sin^4\psi(f^2+4\cos^4\psi) (1+r^4)+(\pi(\nu-1)
+2\psi-\frac{1}{2}\sin4\psi)^2}\,\,\, \times\nonumber\\
\sqrt{1+h(r)\left(r\frac{d\psi}{dr}\right)^2},
\label{routhian7}
\end{align}
where we have rescaled $r$ in the same way as before $r\rightarrow r \sqrt{b}$
and where $\nu$ is defined in equation (\ref{definenu}).
From the Routhian we derive the following equation of motion for $\psi(r)$
\begin{align}
0=&\frac{h\left(r\frac{d}{dr}\right)^2\psi}{1+h\left(r\frac{d\psi}{dr}\right)^2}
-2\frac{r_h^4}{r^4} r\frac{d}{dr}\psi\left[1+\frac{1}{1+h\left({r\frac{d}{dr}\psi}\right)^2}\right]
\nonumber \\
&+hr\frac{d\psi}{dr}\left[ 1+\frac{8r^4\sin^4\psi (f^2+4\cos^4\psi) }{4\sin^4\psi
(f^2+4\cos^4\psi) (1+r^4)+(\pi(\nu-1) +2\psi-\frac{1}{2}\sin4\psi)^2}\right]  \nonumber \\
&-\frac{ 8\sin^3\psi\cos\psi f^2(1+r^4) + 4\sin^32\psi\cos2\psi r^4
+4\sin^22\psi(\pi(\nu-1)+2\psi)
}{4\sin^4\psi(f^2+4\cos^4\psi) (1+r^4) +(\pi(\nu-1) +2\psi-\frac{1}{2}\sin4\psi)^2}.
\label{equationforpsi7}\end{align}
The main difference between the Routhian for the D5 and D7 branes is the term arising from
the charge density, it is $(\pi\nu)^2$ for the D5 brane and $(\pi(\nu-1) +2\psi-\frac{1}{2}\sin4\psi)^2$
for the D7 brane. This difference comes from the presence of Wess-Zumino terms in the action for the
D7 brane.

\section{Characteristics of solutions}
\subsection{Asymptotic behaviour as $r\rightarrow \infty$ \label{Infinity}}
Looking at the equations of motions for $\psi(r)$ for the $D5$
branes and the D7 branes respectively, i.e.\ eqns.\ (\ref{equationforpsi5})
and (\ref{equationforpsi7}), one can
check that the asymptotic behaviour $\psi(r)\rightarrow
\frac{\pi}{2}$ as $r\rightarrow \infty$, assumed in their derivation, is indeed compatible with these.
Expanding $\psi(r)=\frac{\pi}{2}+\delta \psi$ for large $r$ one
furthermore finds the
following differential equation both for D5 and D7
 \begin{align}
\left(r\frac{d}{dr}\right)^2\delta \psi + 3\left(r\frac{d}{dr}\right)\delta \psi + 2\delta\psi=0.
\end{align}
This equation has the  solution $\delta\psi(r)= \frac{c_1}{r}+\frac{c_2}{r^2}$
and hence for large $r$
\begin{align}
\psi(r)=\frac{\pi}{2}+ \frac{c_1}{r}+\frac{c_2}{r^2}+\ldots.
\label{asymptoticpsi}
\end{align}
Since the full equations of motions for $\psi(r)$ are second order differential
equations the two integration constants $c_1$ and $c_2$ completely characterize
the solution. In the dual field theory
language $c_1$ is a quantity proportional to the bare mass of the fundamental
representation fields and $c_2$ is proportional to the chiral
condensate. In the present paper we will always
be dealing with the mass less case, i.e.\ $c_1=0$.

 It is easy to see that the constant function $\psi(r)=\frac{\pi}{2}$ is  a
solution of the equations of motion for all $r\in [r_h,\infty]$ both for
the D5 brane and the D7 brane case. For zero temperature, $r_h=0$,
one can show that
there is a certain critical value of $\nu/f$ below which the constant solution
is unstable, more precisely\footnote{Note that even though the D7 brane
equations of motion depend on $\nu$ and $f$ separately, the prediction
for the location of the phase transition depends only on their ratio.}
\begin{align}
\left(\frac{\nu}{f}\right)_{crit}=\frac{2 \sqrt{7}}{\pi}\approx 1.68,
\hspace{0.5cm}\mbox{for}\hspace{0.5cm} r_h=0.
\end{align}
For $(\nu/f)<(\nu/f)_{crit}$ the stable solution of the equations of motion
should hence be an $r$-dependent solution. When $r_h>0$ we expect that
$(\nu/f)_{crit}$ becomes smaller. A solution with $c_1=c_2=0$ must necessarily
be the constant solution. A solution with $c_1=0$ and $c_2\neq 0$ can be
viewed as showing spontaneous breaking of chiral symmetry. The phase transition
 which occurs when $(\nu/f)=(\nu/f)_{crit}$ is thus interpreted as a
chiral symmetry
breaking/restoring phase transition.
This phase transition was found for the D5 brane
in reference \cite{Jensen:2010ga} and was shown to exhibit
Berezinsky-Kosterlitz-Thouless scaling. For the D7 case numerical
investigations have shown that there are $r$-dependent solutions even in
some part of the region where the constant solution is supposed
to be stable and that
these solution are energetically favoured compared to the constant
one~\cite{Kristjansen:2012ny}.

Finally, let us highlight that the Routhians become identical for the D5 branes
and the D7 brane as $r\rightarrow \infty$ due to the identity
\begin{align}\label{normalizationidentity}
{\mathcal N}_7 = \frac{{\mathcal N}_5 N_5}{f}.
\end{align}

\subsection{Asymptotic behaviour as $r\rightarrow r_h$ \label{horizon}}

Let us consider first the zero temperature case, $r_h=0$.
Looking at the equation of motion for $\psi(r)$ for the D5 brane we see
that for $r=r_h=0$, the equation of motion for $\psi(r)$ reduces to
\begin{align}
\left.\frac{\partial_{\psi} V_5}{2 V_5}\right|_{r=r_h=0}=0,
\end{align}
where the ``potential''
$V_5$ is given by
\begin{align}\label{psipotential}
V_5= 4 \sin ^4\psi f^2(1+r^4) +(\pi \nu)^2.
\end{align}
The angle $\psi$ must hence come to an extremum of the potential $V_5$, i.e.
we need that
\begin{align}
\left.\partial_{\psi} V_5\right|_{r=r_h=0}=
16 \sin^3 \psi \cos \psi f^2=0.
\end{align}
There are only two possible solutions, $\psi=0$ and $\psi=\frac{\pi}{2}$. The
solution $\psi=0$ corresponds to a minimum of
the potential and the solution $\psi=\frac{\pi}{2}$ to a maximum.
A minimum is preferred since the second derivative of is positive there.
When $r_h=0$, if we assume that $\psi\to\psi_0$ as $r\to 0$, the linearized equation in the vicinity of $r=0$ is
$$
\left( r\frac{d}{dr}\right)^2\delta\psi + r\frac{d}{dr}\delta\psi-\frac{\partial^2_\psi V_5(\psi_0)}{2V_5(\psi_0)}\delta\psi=0
$$
and the solution has the form
$$
 \psi \sim \psi_0+\alpha_+ r^{-\frac{1}{2}\left[1+\sqrt{1+\frac{2\partial^2_\psi V}{V}}\right]} +
\alpha_- r^{-\frac{1}{2}\left[1-\sqrt{1+\frac{2\partial^2_\psi V}{V}}\right]}+\ldots
$$
If $\psi_0$ is at a maximum of the potential, so that $\frac{2\partial^2_\psi V_5(\psi_0)}{V_5(\psi_0)}<0$,
both exponents are negative or complex. To have a sensible
solution, both $\alpha_+$ and $\alpha_-$ must be zero.  This means that, if we begin integrating the
nonlinear ordinary differential equation for $\psi(r)$ from $r=0$, the solution will be the constant, and this
can only make sense if $\psi_0=\frac{\pi}{2}$, which is the solution that we already know about.  Coincidentally,
$\frac{\partial^2_\psi V_5(\pi/2)}{V_5(\pi/2)}<0$, so this is a consistent picture.
On the other hand, if $\psi_0$ is a minimum of the potential,  $\frac{\partial^2_\psi V_5(\psi_0)}{4V_5(\psi_0)}>0$,
and one exponent, $\alpha_+$ is negative whereas the other $\alpha_-$ is positive.   To have a sensible
solution,   $\alpha_+$ must be zero. If, again we integrate the differential equation for $\psi$ up from $r=0$,
some fixed value of $\alpha_-$ will lead to an asymptotic form (\ref{asymptoticpsi}) of $\psi(r)$ where both $c_1$ and $c_2$ are nonzero.
We would then have to adjust $\alpha_-$ so that $c_1=0$ to find the type of solutions that we are discussing.  Then $c_2$
is completely fixed by the solution of the equation.
There is the third possibility that the potential is flat, $\frac{\partial^2_\psi V_5(\pi/2)}{V_5(\pi/2)}=0$, which in fact happens at
the other extremum, $\psi_0=0$.  Then $\alpha_+$ must again be set to zero. But the exponent multiplying $\alpha_-$
vanishes and it would seem that the linearized equation is solved by any constant fluctuation of $\psi$.  In this case, one must
appeal to nonlinear effects to see that the correct choice of minimum is still $\psi=0$, although the flatness of the potential in
the vicinity leads to a very slow evolution of $\psi$ toward $\psi=0$ as $r\to 0$.
A similar argument to the above applied to the $r\to\infty$ regime shows that there exist two normalizable modes of the
equation for fluctuations of the angle $\psi$ only when $\psi$ approaches a maximum of the potential, which is the large
$r$ limit of (\ref{psipotential}). The unique maximum is $\psi=\frac{\pi}{2}$ which is the asymptotic value that we are using.

For the D7 brane similar considerations apply but the relevant potential
is different. More precisely,
\begin{align}
V_7=4\sin^4\psi(f^2+4\cos^4\psi)(1+r^4) +(\pi(\nu-1)
+2\psi-\frac{1}{2}\sin4\psi)^2,
\end{align}
for which
\begin{align}
\label{V7prime}
\left.\partial_{\psi} V_7\right|_{r=r_h=0}=
8 \sin ^2\psi \cos \psi \left(2 f^2  \sin \psi +
4   (\pi  (\nu-1)+2 \psi) \cos \psi  \right).
\end{align}
We observe that as in the D5 brane case the derivative of the potential
vanishes for $\psi=0$ and $\psi=\frac{\pi}{2}$. However, in this case there
is a third zero of the derivative which
satisfies
\begin{align}
\frac{f^2 }{2}\tan\psi_0 + \pi(\nu-1)+2\psi_0=0.
\label{psi0}\end{align}
As long as $0<\psi_0<\frac{\pi}{2}$, $\partial_\psi^2V(\psi_0)>0$ and this
point is the minimum of the potential.  
For $\nu<1$ there is always a solution to (\ref{psi0}) in the interval $[0,\tfrac{\pi}{2}]$
but for $\nu>1$ there is never such a solution.
In summary,  for $\nu<1$, $\psi_0$ is always
the minimum. When $\nu>1$, the minimum is at the extreme
point of the interval, $\psi=0$.  Another way to see this is by
looking at  higher derivatives of
the potential. We find
\begin{align}
\left.\partial^2_\psi V_7(\psi)\right|_{\psi=0}=0, \hspace{0.5cm}
\left.\partial^3_\psi V_7(\psi)\right|_{\psi=0}=\frac{32}{3}(\nu-1)\pi.
\end{align}
The vanishing second derivative implies that $\psi=0$ is an inflection point.
Note that the sign of the third derivative is different if $\nu$ is greater or less than one.
If $\nu>1$, the potential is decreasing as $\psi$ approaches zero and the endpoint of
the interval is  a global minimum for the function restricted to the range $[0,\tfrac{\pi}{2}]$.
On the other hand,  if $\nu<1$ it is increasing as $\psi$ approaches zero, and the endpoint
is a local maximum, the only minimum being at $\psi=\psi_0$.

The above applies for when $\nu>1$ or $\nu<1$.  However, for the D7 brane, $\nu=1$
is a special place.
For $\nu=1$ the potential $V_7(\psi)$ vanishes for $\psi=0$ and the last term
in eqn.~(\ref{equationforpsi7}) diverges.  The equation
can still be fulfilled if $\psi$ becomes zero at some value of $r=r_0>r_h$
and if $d\psi/dr$ simultaneously diverges at  $r_0$. This type of embedding of
the D7 brane is known as a Minkowski embedding, the D7 brane pinches off at AdS radius
$r_0$ and does not reach the horizon. It is not possible to have
a Minkowski embedding for the D7 brane for other values of $\nu$. For the
D5 brane a Minkowski embedding would only be possible for vanishing charge
density, i.e.\ for $\nu=0$, a case we shall not be interested in here.

For $r_h\neq 0$, the equation of motion evaluated at $r=r_h$ is
\begin{align}
0=
-4  r_h\frac{d}{dr}\psi
+\frac{\partial_\psi V(\psi)}{2V(\psi)}
\label{equationforpsi7atrh}\end{align}
This no longer requires that $\psi$ goes to an extremum of the potential, but it
determines the derivative of $\psi$ at the horizon once a value of $\psi$ is specified
there.

\subsection{Composite systems}

As explained above the parameters of our  $N_5$ probe D5 branes
and our single probe D7 brane are adjusted
in a particular way in order to allow the interpretation of the probe D7 brane
as a blown up version of the $N_5$ D5 branes. More precisely we fix the flux
through the extra 2-sphere wrapped by the D7 brane to fulfill the
relation~(\ref{definef}) and we adjust the charge density and the
magnetic field so that it is the
same for the D5 branes and the D7 brane. Starting from $N_5$ D5 branes with
a given charge density one can, however, imagine other scenarios than all of
them blowing up to a single D7 brane.

For instance $n_5$ D5 branes could blow up to a D7 brane and the rest remain
D5 branes. The
resulting brane configuration in the interior of $AdS_5$ would then be
a single D7 brane with flux $f=\frac{2\pi}{\sqrt{\lambda}}n_5$ and
$(N_5-n_5)$ D5 branes. The charge density would have to
be shared between the D5 branes which would remain D5 branes and those
which would blow
up, resulting in, for instance for initial filling fraction $\nu$,
the D7 branes having filling fraction $\nu-\nu_0$ and the
remaining $D5$ branes having filling fraction $\nu_0$.

Similarly, the $N_5$ D5 branes could blow up to a larger number of D7
branes with different charge densities. In the most general case, starting
from $N_5$ D5 branes with filling fraction $\nu$ we could have $N_5-n_5$ D5
branes remaining D5 branes with filling fraction $\nu_0$
and
$n_5$ D5 branes blowing up to $n_7$
D7 branes with flux values  $\{f_i\}_{i=1}^{n_7}$,
and filling fractions $\{\nu_i\}_{i=1}^{n_7}$
where the parameters would have to fulfill
\begin{align}
\nu=\nu_0+\sum_{i=1}^{n_7}\nu_i,
\hspace{0.5cm} n_5=\frac{\sqrt{\lambda}}{2\pi}\sum_{i=1}^{n_7}f_i.
\end{align}
Notice in particular that this implies that the Routhian of the composite
system (assuming the components to be non-interacting) becomes identical
to the simple D5 brane Routhian as $ r\rightarrow \infty$.

We shall not study this most general composite system but limit ourselves
to the case where $\nu_i=1$, for all $i$, except possibly for one,
and $0\leq \nu_0<1$. The reason for this
is that the $\nu=1$ gapped D7 brane has a special status, being particularly
favoured energetically and according to our previous
studies~\cite{Kristjansen:2012ny} having the interpretation of a first
quantized Hall level. Composite systems will be investigated in detail
in section~\ref{composite}.
.
\section{Numerical Investigations}

We wish first in subsection~\ref{characteristic}-\ref{crossover} to
consider a situation where a single D7 brane can be
viewed as a blown up version of  $N_5$ D5 branes. Accordingly,
 we chose the same
value of the B-field, the charge density and hence the filling fraction
$\nu$ for the two systems. Obviously, $r_h$ is also chosen to be the same
for the two systems.
Furthermore, we choose the flux of the D7 brane
on the second two-sphere, $\tilde{S}^2$, to be given as
$f=\frac{2\pi}{\lambda}N_5$. Subsequently, in subsection~\ref{composite}
we turn to considering the types of composite systems mentioned above.
I all cases we will restrict ourselves to the massless case, i.e.\ $c_1=0$,
cf.\ section~\ref{Infinity}.

\subsection{Characteristic solutions \label{characteristic}}

To generate a non-constant solution for $\nu\neq 1$ we generate from the differential equation a Taylor series
for $\psi$ as a function of $r$ for $r$ in the vicinity of $r_h$ assuming some value $\psi_0$ for $\psi(r_h)$. This
series expansion is then used to generate the initial data needed for the integration procedure. We finally
determine the value of  $\psi_0$ by demanding
that the solution has $c_1=0$.

\begin{figure}
\begin{center}
\includegraphics[scale=.5]{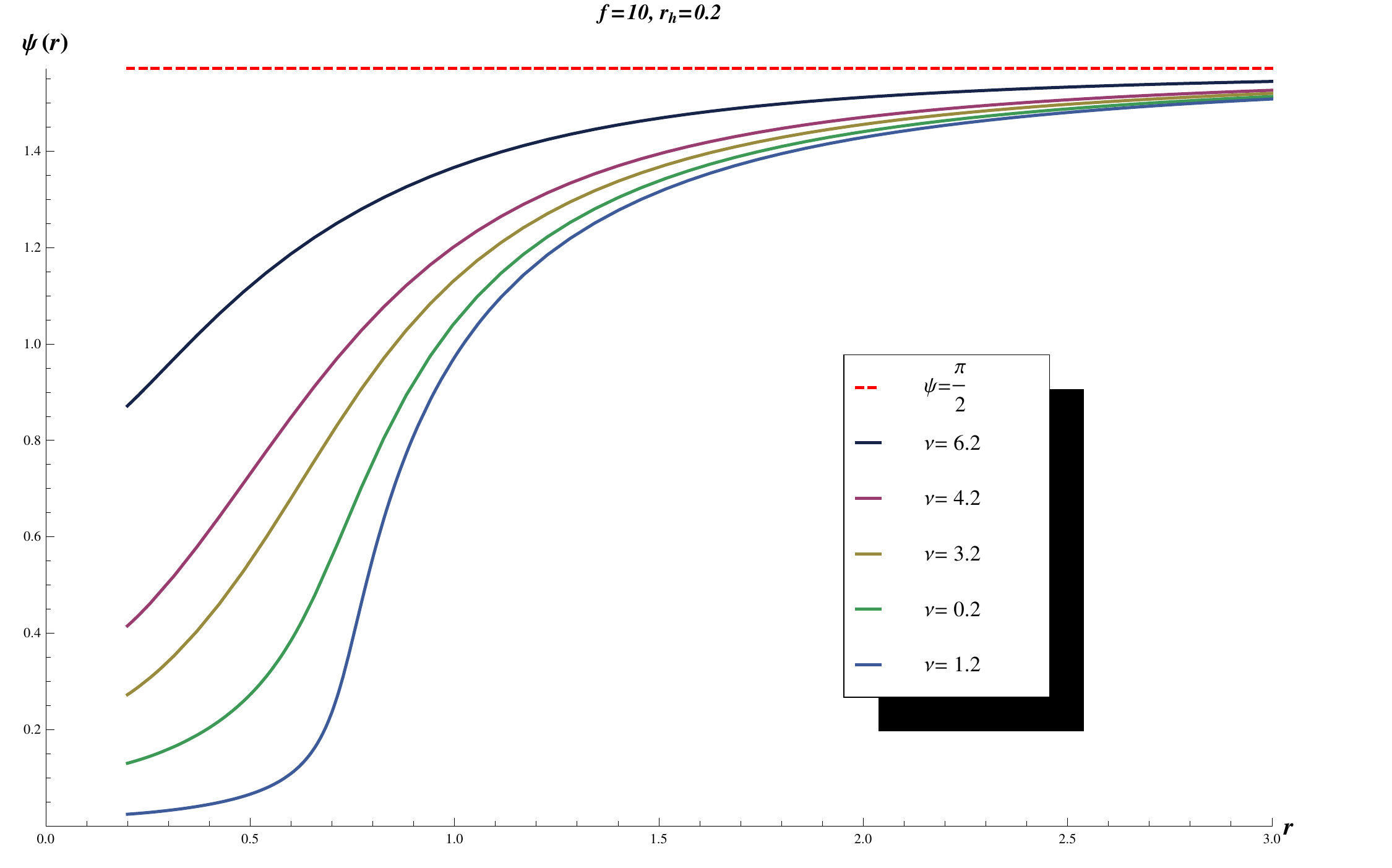}
\end{center}
\caption{\label{charsol} D7 brane solutions for $f=10$ and $r_h=0.2$  for
various values of $\nu$.}
\end{figure}

To generate a non-constant solution for $\nu=1$ we generate
from the differential equation a Taylor series solution for  $r(\psi)$ for
small $\psi$ assuming some value $r_0$ for $r(\psi=0)$. Then we use
 the value of $r(\psi)$ and $r'(\psi)$ as generated by this
Taylor series expansion as input to our integration procedure. Analogously to above
we determine the value of $r_0$ by demanding that the solution has $c_1=0$.
In figure~\ref{charsol} we have plotted some  solutions for $\psi(r)$
for $f=10$ and $r_h=0.2$ and various values of $\nu$.

\subsection{The stability lines for D5 and D7 \label{stability}}

As explained in subsection~\ref{Infinity} the constant solution
$\psi=\frac{\pi}{2}$ solves the equation of motion both for D5 branes and
D7 branes
for all values of the
parameters but this solution is expected to be unstable when $\nu/f$ is small.
For $r_h=0$  the critical point below which the constant solution is unstable
is given by  $(\nu/f)_{crit}=\frac{2\sqrt{7}}{\pi}$.
In this case, however, numerical studies show, that the non-constant D7
brane
 solution remains energetically favoured even in a part of the parameter space
where the constant solution is stable~\cite{Kristjansen:2012ny}.

One would expect that also for $r_h\neq 0$ one would have a region for small
values of $\nu/f$ where the non-constant solution is energetically favoured.
We have investigated this by comparing the energies of constant and
non-constant solutions as determined from (minus) the corresponding values
of the Routhian for various values of our parameters. Notice that whereas
the total energy of any of the systems considered diverges
(cf.\ eqns.~(\ref{Routhian5}) and~(\ref{routhian7})) energy differences between
systems with identical values of the parameter $f$ are finite.

In figure~\ref{figure003} we have shown the transition lines separating the
region where the non-constant solution is energetically favoured from the
region where the constant solution is the favoured one.
The red curves correspond to D5 branes and the blue ones to D7 branes.
Notice that non-constant D5 brane solutions depend only on $\nu/f$ whereas
non-constant D7 brane solutions depend on $\nu$ and $f$ separately.
(The phase diagram for the D5 brane appeared already in \cite{Evans:2010hi}.)

As for the zero-temperature case the D7 brane seems to have a much bigger
region where the non-constant solution is favoured and very likely the
non-constant solution  again co-exists with the stable constant solution in
a large part of the parameter space. The end point of the transition lines
at $r_h=0$ coincide with our previous zero-temperature
estimates~\cite{Kristjansen:2012ny}.

It is interesting to notice that the plots all have a peak corresponding
to $\nu=1$ which shows that this value of the filling fraction is particularly
favoured. This is in agreement with our earlier interpretation of this state
as the first quantum
Hall level~\cite{Kristjansen:2012ny}.
 The special status of the $\nu=1$ state implies that
it is advantageous for the D branes to
organize into composite systems for $\nu>1$. We shall discuss this in
detail in section~\ref{composite}.

\subsection{Crossover between D5 and D7 for $\nu<1$ \label{crossover}}

To the left of the blue curves in  figure~\ref{figure003}
the D5-branes and the corresponding D7 brane both have lower energy than
the constant solution. It is thus interesting to investigate which one of these
two has the lowest energy in the region $0<\nu<1$. (As already
mentioned, when we pass the line $\nu=1$, we in addition
have composite systems to worry about
and this case will be discussed in the following subsection.) We have earlier
pointed out that the Routhians for the D5-branes and the D7 branes become
identical when $r\rightarrow \infty$. We notice that for $r\rightarrow r_h$
the Routhians would coincide for $\nu=1/2$ if for both systems the angle
$\psi$ would tend to zero at the horizon and a reasonable first guess for
the location of the transition point could be at $\nu=1/2$.
(We know, however, that for the D7 brane when $r_h\neq 0$  
and $\nu< 1$ the angle $\psi$
does not tend to zero at the horizon, cf.\ section \ref{horizon}.)

In figure~\ref{figure004} we show in green the line of transition between D5
and D7 for $0<\nu<1$ for various values of $f$.
The curve lies somewhat displaced from $\nu=\frac{1}{2}$ but
approaches this line when $f$ becomes larger.

\subsection{Composite systems \label{composite}}

As discussed above the gapped D7 brane with $\nu=1$ is particularly
energetically
favoured. One can hence wonder whether composite systems could start playing
a role when $\nu>1$. Let us consider $\nu=1+\nu_0$, where $\nu_0<1$. For this
value of $\nu$ one could imagine that $n_5$ of the
$N_5$ D5 branes would blow up to a D7 brane with $\nu=1$ and the
rest either remain D5 branes with $\nu=\nu_0$
or blow up to another D7 brane with $\nu=\nu_0$. To see if this possibility
is realized we have to compare the energy of the composite system with that
of the simple D7 brane and D5 brane solution. The energy of the
composite solution will of course depend on
how many D5 branes blow up to gapped D7 branes and how many remain ungapped
branes. The distribution of the D5 branes is reflected in the parameter
$f$ of the two components of the composite system. Let us denote the
the flux of the gapped brane as $f_0$, i.e.\
\begin{align}
f_0=\frac{2\pi}{\sqrt{\lambda}}n_5.
\end{align}
Then the $f$-parameter of the ungapped branes becomes $f_{tot}-f_0$ where
$f_{tot}$ is the $f$-parameter of the initial D5 branes.
What we are interested in is the composite system for which the energy is
the smallest possible one so for a given initial number $N_5$ of D5 branes
and hence a given initial value of $f_{tot}=\frac{2\pi}{\sqrt{\lambda}}N_5$
we will have to  find the value of $f_0$  which
minimizes the energy of the  composite systems. Naively one would expect
that as many D5 branes as possible would blow up to gapped D7 branes but
there are many dynamical issues which must be taken into account and we
have to determine the minimum energy solution numerically.

\begin{figure}[h]
\begin{center}
\includegraphics[width=7.0cm]{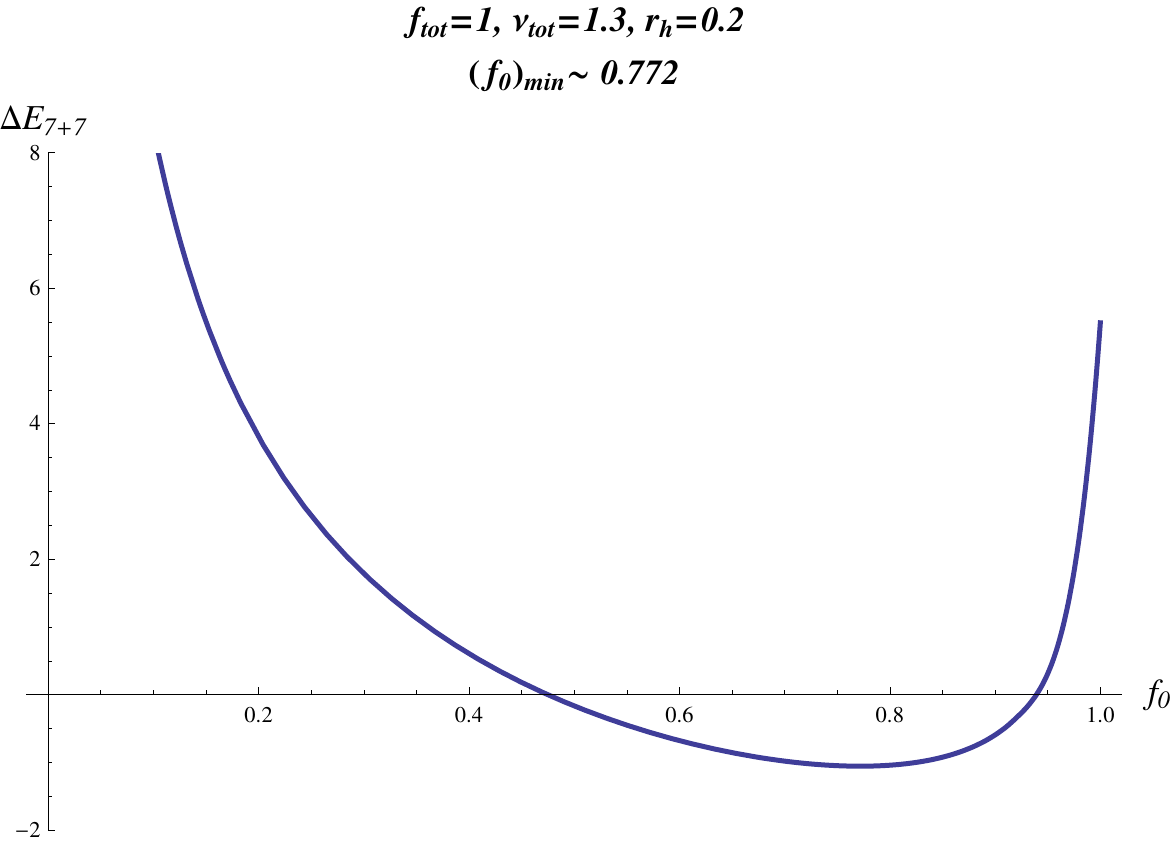}
\end{center}
\caption{\label{Minimization}Plot for $f_{tot}=1$ and $r_h=0.2$
showing the energy of
a composite system consisting consisting of a gapped i.e.\ $\nu=1$ D7 brane with
flux $f_0$ and  ungapped D7 branes with $\nu=0.3$ and
$f=1-f_0$  minus the energy of the constant solution with $\nu=1.3$.
The energetically favoured solution corresponds to $f_0=0.772$.}
\end{figure}

In figure~\ref{Minimization} we show an example for $f_{tot}=1$ and $r_h=0.2$
of how we sweep over different values of $f_0$ to determine the minimum
possible energy for the composite system. Here we are considering a
composite system  consisting of a
gapped D7 brane with $\nu =1$ and a number of un-gapped D7 branes with
$\nu=0.3$. A similar sweep over values of $f_0$ must be done for the competing
system consisting of gapped D7 branes supplemented with D5 branes.
Subsequently, we can compare the minimum energies of the the two composite
systems and in addition we should compare these to the energy of the
non-constant D7-brane solution with $\nu=1.3$ and $f=1$. (Had there been
a non-constant D5 brane solution with similar parameters we would also
have had to compare to the energy of this one but there is not.) In this way,
i.e. by comparing energies, we are able to tell which system is the favoured
one.

The case we have discussed pertains to the situation $1<\nu<2$. Let us now
discuss what happens when we vary $\nu$ in this range. What we find is that
when $f$ is small the composite system consisting of gapped D7 branes plus
un-gapped D7 branes is always the favoured one. However when $f$ becomes
larger, there appears at a certain value of
$\nu$ a  crossover between a region where the favoured composite system
is D7 plus D5 and a region where the favoured composite system is D7 plus
D7.
\begin{figure}[h]
\begin{center}
\includegraphics[scale=0.65]{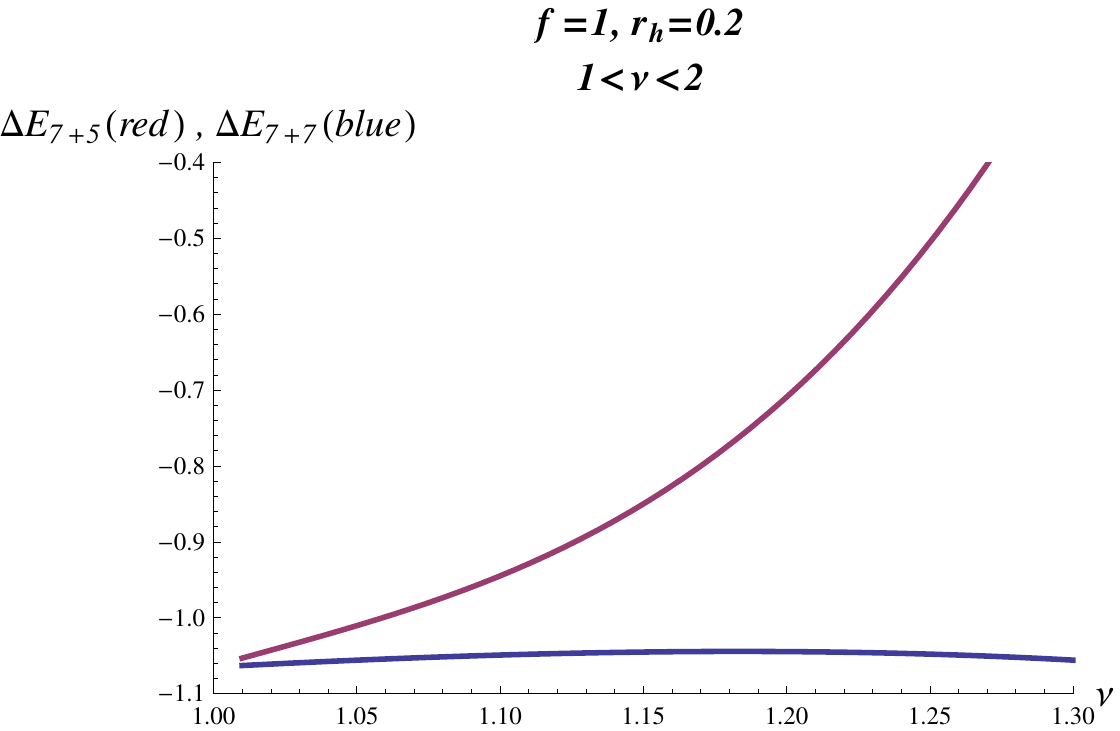}
\includegraphics[scale=0.65]{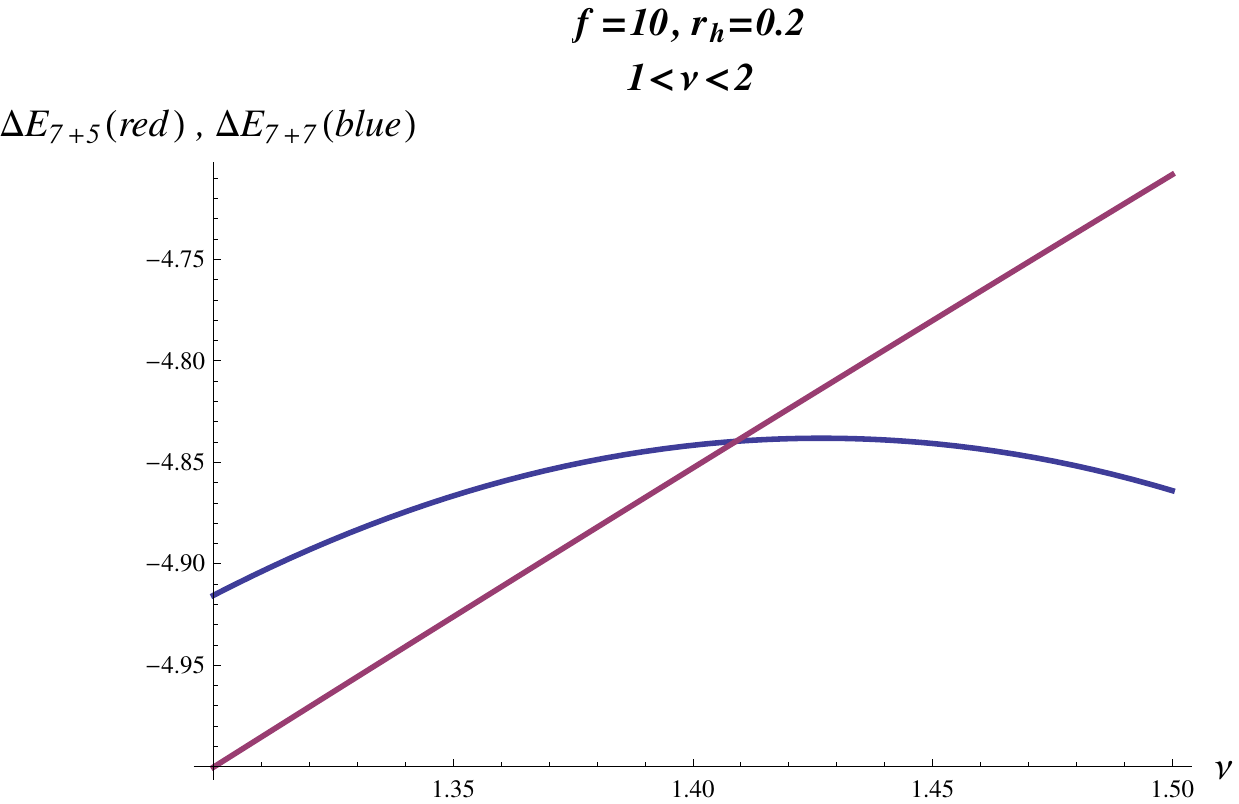}
\end{center}
\caption{\label{compositecompetition} The difference between the
energy of the composite D7-D5 system and the constant solution
(red curves) and the difference between the energy of the
composite D7-D7 system and the constant solution (blue curves) for
$r_h=0.2$ and  $f=1$ and $f=10$ respectively. Notice the crossover
at $\nu\approx 1.41$ for $f=10$. }
\end{figure}
In figure~\ref{compositecompetition} we show for  $r_h=0.2$ and
$f_{tot}=1$ the energy difference between the D7-D5 system and the
constant solution (red curves) and the energy difference between
the D7-D7 brane system and the constant solution (blue curves) for
$f_{tot}=1$ and for $f_{tot}=10$ as a function of $\nu$ where
$\nu\in [1,2]$. Notice that to generate a given data point on each
of these curves we first have to go through the minimization
procedure described above and illustrated in
figure~\ref{Minimization}. The curves tell us that for $f_{tot}=1$
and $r_h=0.2$ the composite D7-D7 system is always the favoured
one but for $f_{tot}=10$ and $r_h=0.2$ there is a crossover
between D5-D7 and D7-D7 at $\nu\approx 1.41$. We have repeated the
analysis for different values of $r_h$ and found that the
crossover point does not show strong dependence on $r_h$.

Now we can move on to considering the interval $2<\nu<3$, i.e. a $\nu$ on
the form $\nu=2+\nu_0$ where $0<\nu_0<1$.
In this interval we can have composite systems consisting of two
gapped D7 branes with $\nu=1$ in combination with either ungapped D7 branes
or D5 branes. Again we have to determine by numerical investigations how
many D5 branes blow up to gapped D7 branes and how many do not. In addition,
we now in principle have the option that the two gapped D7 branes can have
different values for the flux. However, we know from our previous analysis
of the zero temperature case~\cite{Kristjansen:2012ny} that for a collection
of gapped D7 branes with total flux $f$ the energetically favoured situation
is the one where the flux is equally shared between the D7 branes. If we
denote the flux for each of the gapped D7 branes as $f_0$ the $f$-parameter
for the un-gapped branes now becomes $f_{tot}-2 f_0$. Again we have to sweep over
$f_0$ to determine how precisely
the branes of the composite systems organize themselves
into gapped branes and ungapped ones. After having found the most favourable
configuration for each of the two types of composite systems we can again
compare their energies to each other and to the energy of the constant
solution with $\nu=2+\nu_0$. What we find is that the pattern seen in the
interval $1<\nu<2$ repeats itself. For small values of $f$ the composite
D7-D7 system always wins but when $f$ becomes larger there starts to appear
a cross over between D7-D5 and D7-D7. Again the cross over point does not
depend very much on $r_h$.

\begin{figure}[h]
\begin{center}
\includegraphics[scale=0.7]{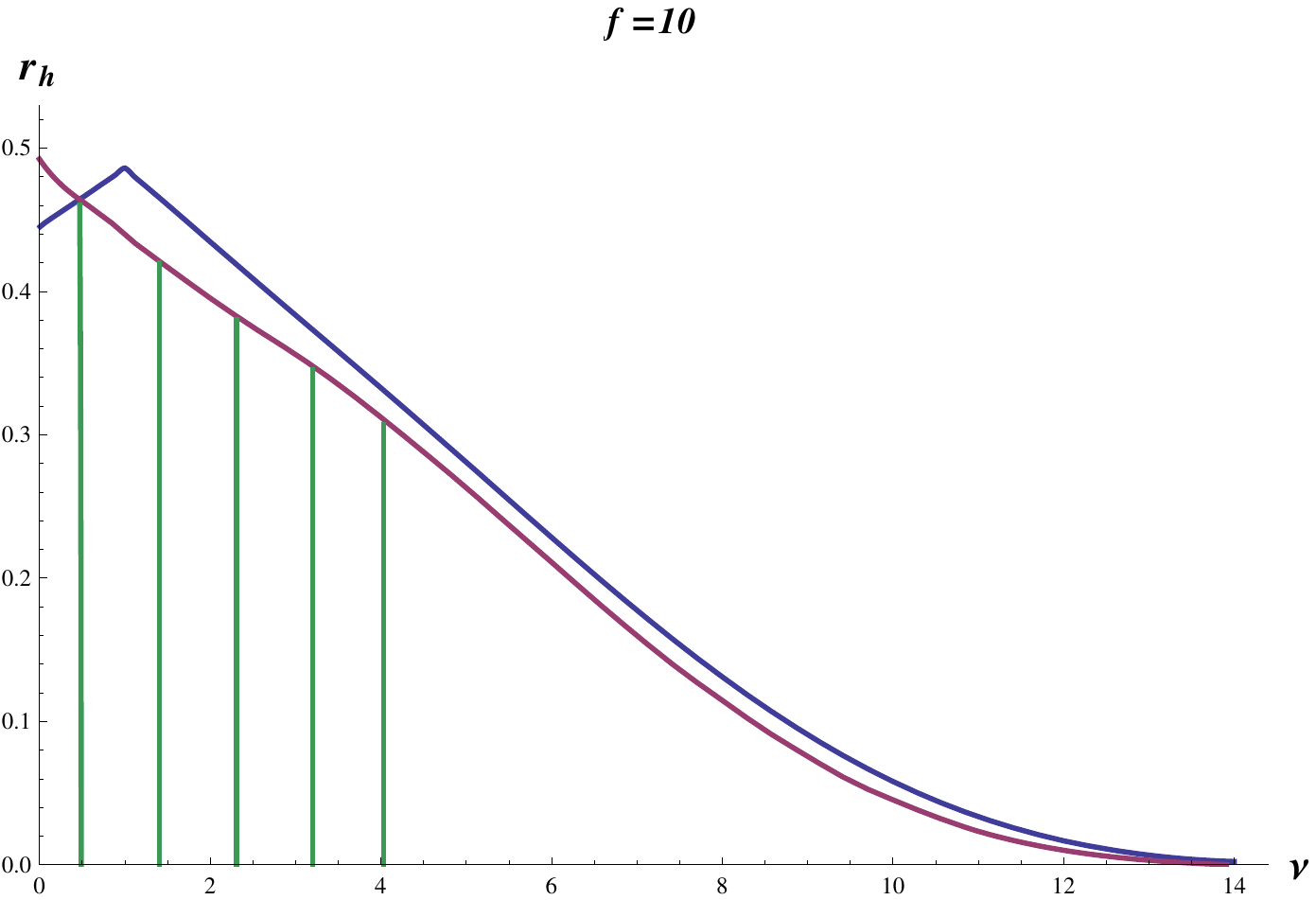}
\end{center}
\caption{\label{Compositeplot}Plot for $f=10$ showing a transition line
(green) in each interval $[\nu,\nu+1]$ with $\nu\in \{0,1,2,3,4\}$ separating the composite D5-D7
system with $\nu$ gapped D7 branes from the composite D7-D7 system, likewise
with $\nu$ gapped D7 branes. For $\nu>5$, the D7-D7 system always wins.}
\end{figure}

 It is obvious that we can now repeat the whole procedure again in the
interval $3<\nu<4$ and in all the following intervals of the type
$n<\nu<n+1$ where we could have composite systems with $n$ gapped
D7 branes in combination with an ungapped D7 brane or with D5 branes.
We have done the analysis for $r_h=0.2$ and for intervals up to and including $\nu\in[8, 9]$.
We have found that up to and including  $\nu\in [4,5]$ there
is in each interval a transition between a region where the D7-D5 system is the
energetically favoured one and another region where the D7-D7 system
is favoured. The region where the D7-D5 system is favoured diminishes as
$\nu$ increases and
for $\nu>5$ the D7-D7 system  always wins.
In figure~\ref{Compositeplot} we show the full phase diagram for $f=10$.
The red and the blue curves are the stability lines for the D5 and the D7
branes from figure~\ref{figure003} and the vertical green lines are the
lines which separate the composite D5-D7 systems from the composite D7-D7
systems.

\section{Conclusion}

The results of our numerical investigations and the conclusions that can be drawn from there are reviewed
in the first section of this paper.   In this concluding section, we wish to point out some interesting directions for
further work on this subject, including some speculations about possible new results.

We have not explored the blown up solutions of the D5 brane from the D5 brane point of view where it would be a non-Abelian configuration
of D5 branes.  There are a number of obstacles to this approach, one being that the full generalization of the Born-Infeld action is not known when the
embedding coordinates of the D brane are matrices.   It would nevertheless be interesting to ask whether
some of the features of the solution that we find are visible in the non-Abelian D5 brane theory.  We expect that the approximation of the non-Abelian
D5 brane as a classical D7 brane is good in the limit where the number of D5 branes is large.   However, we also expect that the blow up
phenomenon at $\nu=1$ should also be there for a small number of branes and the only way to see it is from the non-Abelian D5 brane point of view.

We have done extensive numerical solutions of the embedding equations for the D5 and D7 branes.  We have not analyzed the small
fluctuations about these solutions.   The spectrum of fluctuations would tell us, for example, if the solutions that we have found are stable
or metastable.   A search for further instabilities would be very interesting, especially considering that other D7 brane configurations are
known to have instabilities to forming spatially periodic structures when the density is large enough.

Most excitingly, we have an observation about a possibly interesting electronic property of the ungapped D7 brane solutions.
If we compare the numerical solutions of the ungapped Abelian D5 brane and the ungapped D7 brane, in the
regions of the phase diagram where the D7 brane is favored, it has lower energy because it has an exceedingly narrow funnel.
The funnel is the lower AdS radius part of the D brane world volume which is in the region where the brane approaches the Poincare horizon.
Numerical computation in reference  \cite{Kristjansen:2012ny} already found this narrow funnel and we have confirmed it here.
As we have discussed in section 1, when the
filling fraction is not an integer, it is necessary for the D brane worldvolume to reach the Poincare horizon.  This is due to the fact that
the worldvolume electric fields that are needed to create the nonzero charge density state in the holographic theory need to end at a source, or else
thread through the entire space.  In a rough sense, the D brane creates the source in that it becomes narrow to emulate a group of suspended strings.
In the case of the D7 brane, this funnel region is exceedingly narrow.  In the dual field theory, this narrowness of the D brane funnel
implies a very small density of electronic states at low energies.  We liken this to the situation at weak coupling where localization due to impurities depletes
the extended states, leading to a mobility gap.  For the D7 brane, there is no mobility gap in the mathematical sense,
but the density of conducting states is anomalously
small.  This scarcity of conducting states could lead to an approximate, dynamically generated quantum Hall plateau,
a tantalizing possibility since  the only other known
mechanism for Hall plateaus is localization.   This would give a strong coupling mechanism.

We have not found solutions of probe D5-brane theory which would be the holographic dual of fractional quantum Hall states.  
Such states are both predicted by theoretical arguments \cite{frac1}-\cite{frac6} and found experimentally 
\cite{frac7}-\cite{frac9} in graphene and
they are normally taken as evidence of strong electron-electron
correlations.  The strong coupling limit that we can analyze using holography
should be expected to see such states.   One might speculate that solutions corresponding to fractional Hall states could be obtained
from the integer quantum Hall states that we have already found by SL(2,Z) duality which has a natural realization in three
dimensional Abelian gauge theory \cite{Witten:2003ya}
and  also   a natural action on quantum Hall states \cite{Burgess:2000kj} \cite{Burgess:2006fw}.  
In particular, it can map integer 
quantum Hall states to fractional quantum Hall states.   
Exactly how this would work in the context of the present paper certainly merits further careful study.  
In particular, it could elucidate the relationship of the current work with other known string theory and holographic 
constructions of fractional quantum Hall states \cite{Brodie:2000yz}-\cite{Leigh:2012jv}. 

A beautiful aspect of the fractional Hall states found in reference \cite{Fujita:2009kw} is the explicit construction of boundaries and the
existence of boundary currents.  In our construction, the Hall state has a charge gap and it must therefore be a bulk insulator.
The Hall current should be carried by edge states.  The edges must be at the asymptotic spatial boundaries.  It would be interesting,
following reference \cite{Fujita:2009kw} to attempt to construct boundaries or domain walls which would carry the currents.   

Another place that SL(2,Z) duality and alternative quantization have been exploited recently is in the holographic construction of 
an anyonic superfluid \cite{Jokela:2013hta}.  That construction was based on the D7' model which has non-integer quantized Hall states.  It exploited
the idea that, whcn the external magnetic field is made dynamical, so that it adjusts its own vacuum expectation value to the desired
filling fraction, what was a quantum Hall state obtains a soft mode and becomes a compressible superfluid. 
It should be possible to apply similar reasoning to the construction that we have outlined in this paper.  In this case, it would
describe anyons based on integer level Chern-Simons theory.

\section*{Acknowledgments}
C.\ Kristjansen was supported by FNU through grant number DFF -- 1323 -- 00082.  \mbox{R.\ Pourhasan} and G.\ Semenoff were supported by
NSERC of Canada. R.\  Pourhasan was supported in part by Perimeter Institute for Theoretical Physics. Research at Perimeter Institute is supported by the Government of Canada through Industry Canada and by the Province of Ontario through the Ministry of Research and Innovation. R.\ Pourhasan would like to thank Niels Bohr Institute for their hospitality during the visit where part of this work was done.
 G.\ Semenoff acknowledges the kind hospitality and financial support of the International Institute of Physics in Natal, Brazil, where this work was completed.

\end{document}